\documentclass[12pt]{article}
\usepackage{geometry}                
\usepackage{graphicx,subcaption}
\usepackage{amssymb}
\usepackage{amsmath,color}
\usepackage{tikz}
\usepackage{tabularx}
\usepackage{multirow,ulem}

\newcommand{\keff}{k_\mathrm{eff}}

\newcommand{\mA}{\mathbf{A}}

\newcommand{\qref}[1]{Eq.~\eqref{#1}}
\newcommand{\sigmas}{\sigma_\mathrm{s}}
\newcommand{\sigmat}{\sigma_\mathrm{t}}

\newcommand{\sigmaf}{\sigma_\mathrm{f}}

\newif\ifclean

\cleantrue

\ifclean
\newcommand{\change}[1]{\textcolor{black}{#1}}
\renewcommand{\sout}[1]{}
\else

\newcommand{\change}[1]{\textcolor{magenta}{#1}}
\fi

\title{Calculating Time Eigenvalues of the Neutron Transport Equation with Dynamic Mode Decomposition}
\author{Ryan G.\ McClarren
\\ Dept.\ of Aerospace and Mechanical Engineering\\University of Notre Dame\\365 Fitzpatrick Hall, Notre Dame, IN 46556 USA}

\begin{document}
\maketitle

\begin{abstract}
A novel method to compute time eigenvalues of neutron transport problems is presented based on solutions to the time dependent transport equation. Using these solutions we use the dynamic mode decomposition \change{\sout{(DMD)}} to form an approximate transport operator.  This approximate operator has eigenvalues that \change{\sout{can be directly} are mathematically} related to the time eigenvalues of the neutron transport equation.  This approach works for systems of any level of criticality and does not require the user to have estimates for the eigenvalues.  Numerical results are presented for homogeneous and heterogeneous media.  The numerical results indicate that the method finds the eigenvalues that are \change{\sout{most important to the solution evolution} that contribute the most to the change in the solution}  over a given time range, and the eigenvalue with the largest real part is not necessarily important to the system evolution \change{at short and intermediate times}. 
\end{abstract}

\section{Introduction}
In scientific computing we are used to taking a known operator and making approximations to it. Usually these approximations arise from the continuous operator and restricting it to some discrete representation.   This is what is done in common methods for particle transport such as discrete ordinates where the continuous transport equation is replaced with equations for particular directions that are coupled through scattering via a quadrature rule.

Alternatively, it is possible to use the action of the operator to generate approximations rather than using the operator itself. This is what is done in, for example, Krylov subspace methods for solving linear systems where the action of a matrix is used to create subspaces of increasing size that are used to find approximations to the solution. The use of the known action of an operator, even if the operator is not known, is the basis for the dynamic mode decomposition (DMD) \cite{ROWLEY:2009hb,Schmid:2010ee}.

The main idea behind DMD is that if we have a sequence of vectors generated by successively applying an operator, we can estimate properties of that operator. In fluid dynamics,  DMD is used to find important modes in the evolution of a system, even when the system does not have an interesting steady state \cite{Schmid:2010ee,Schmid:2010hh}.  Additionally, because one does not need the operator, DMD can be applied to experimental measurements and quantitively compared to the DMD modes of a simulation \cite{Schmid:2011ec}. 

In this paper we use DMD to find time eigenvalues, also known as $\alpha$ eigenvalues, of the neutron transport equation using only the time dependent solution for the angular flux.  The calculation of $\alpha$ eigenvalues has traditionally been accomplished using iterative search procedures where an eigenvalue is determined by finding the value of $\alpha$ that makes the equivalent $k$-eigenvalue problem exactly critical \cite{hill:1983wj}.  This is accomplished by subtracting $\alpha$ divided by the neutron speed from the total interaction term.  Unfortunately, if the $\alpha$ eigenvalue is negative (that is, the system is subcritical), a negative total interaction term can result, leading to instabilities in most solution algorithms.  Recently, there have been improvements to deterministic $\alpha$ eigenvalue computation techniques that use specialized solvers to find positive and negative eigenvalues \cite{Ortega:2017tw,Lathouwers:2003fv,Modak:2007hz} \change{or form the full discretization matrices to find eigenvalues \cite{modak2003simple}}.  Most of these methods either find only the eigenvalue with the largest real part (the rightmost eigenvalue in the complex plane), or require an accurate estimate to find other eigenvalues. Additionally, Monte Carlo can be used to find these eigenvalues with the transition rate matrix method \cite{Betzler:2015ey,Betzler:2018dv}.

The benefit of using the DMD method is that one can use standard transport solvers \cite{LewisMiller} to find \change{any eigenvalues that are excited} in a given calculation. The cost of the calculation, beyond the transport simulation, is the formation of a singular value decomposition (SVD) on the solution at several time steps. No development of transport solvers is required and off-the-shelf linear algebra routines can be used to find the SVD. DMD will find the eigenvalues/eigenvectors that are \change{\sout{strongest influences on}  the largest contributors to} the dynamics of the system in a given time dependent problem: this a feature and not a bug.  In many subcritical systems the rightmost eigenvalue will be unimportant to the system behavior in a given experiment.  For instance, if we consider a subcritical system struck by a pulse of neutrons, such as those in \cite{Procassini:2010vn}, there will be eigenmodes corresponding to the slowest neutrons traveling across the system \cite{Larsen:1974hd,Larsen:1979hs} that will not impact the experiment. We will see an example of this later.

This paper is organized as follows. We begin with the presentation of the dynamic mode decomposition in section \ref{sec:dmd}, and apply the method to time-eigenvalue problems in \ref{sec:alphas}.  Numerical results are presented for a bare sphere in section \ref{sec:sphere_results} and for heterogeneous systems in \ref{sec:hetero} before presenting conclusions and future work in section \ref{sec:discussion}.

\section{Dynamic Mode Decomposition}\label{sec:dmd}

Consider an evolution equation over time that can we written in the generic form
\begin{equation}\label{eq:evol}
\frac{\partial y}{\partial t} = A(r) y(r,t),
\end{equation}
where $y(r,t)$ is a function of a set of variables denoted by $r$, which could be space, angle, energy, etc., and time $t$. Consider the solution to the equation at a sequence of equally spaced times, $y(r,t_0), y(r,t_1), \dots, y(r,t_{N-1}), y(r,t_N)$, separated by a time $\Delta t$.  These solutions are formally determined using the exponential of the operator $A(r)$ via the relationship,
\[ y(r,t_n) =  e^{A \Delta t}y(r,t_{n-1}), n= 1, \dots, N.\]
We can write a single equation relating the solutions at each time level as
\begin{equation}\label{eq:expA}
[y(r,t_N), y(r,t_{N-1}), \dots, y(r,t_1)] = e^{A \Delta t} [y(r,t_{N-1}), y(r,t_{N-2}), \dots, y(r,t_0)].
\end{equation}

If we constrain ourselves to finite dimensional problems, the solution is now a vector and the operator is a matrix.  In this case the original equation has the form
\begin{equation}\label{eq:evol_finite}
\frac{\partial \mathbf{y}}{\partial t} = \mathbf{A} \mathbf{y}(t).
\end{equation}
We will say that $\mathbf{y}_n$ is of length $M > N$ and $\mathbf{A}$ is an $M \times M$ matrix.
In this case, the solutions are related through the matrix exponential:
\begin{equation}\label{eq:expAmat}
[\mathbf{y}_N, \mathbf{y}_{N-1}, \dots, \mathbf{y}_1] = e^{\mathbf{A} \Delta t} [\mathbf{y}_{N-1}, \mathbf{y}_{N-2}, \dots, \mathbf{y}_0] .
\end{equation}
In shorthand we can define the $N \times M$ matrices \[ \mathbf{Y}_{+} = [\mathbf{y}_N, \mathbf{y}_{N-1}, \dots, \mathbf{y}_1] , \qquad   \mathbf{Y}_{-} = [\mathbf{y}_{N-1}, \mathbf{y}_{N-2}, \dots, \mathbf{y}_0] ,\] as the matrices formed by appending the column vectors $\mathbf{y}_n$. This leads to the relation
\begin{equation}\label{eq:expAmatU}
\mathbf{Y}_+ = e^{\mathbf{A} \Delta t} \mathbf{Y}_{-} .
\end{equation}

Equation \eqref{eq:expAmatU} is exact; however the matrix $\mA$ may be too large to compute the exponential, $e^{\mathbf{A} \Delta t}$. 
Therefore, we desire to use just the solution to estimate the eigenvalues of $e^{\mathbf{A} \Delta t}$.

%

To this end we will use the solution vectors collected in $\mathbf{Y}_+ $ and $\mathbf{Y}_{-}$ to produce an approximation to $\mA$.  We compute the thin singular-value decomposition (SVD) of the matrix $\mathbf{Y}_{-}$:
\begin{equation}\label{eq:SVDUjm1}
\mathbf{Y}_{-} = \mathbf{U}\boldsymbol{ \Sigma}\mathbf{ V}^*,
\end{equation}
where $\mathbf{U}$ is an $M \times N$ unitary matrix, $\mathbf{V}$ is a $N \times N$ unitary matrix, and $\boldsymbol{\Sigma}$ is an $N \times N$ diagonal matrix with non-negative elements. The asterisk denotes the conjugate-transpose of a matrix. Typically, some of the diagonal elements of $\boldsymbol{\Sigma}$ are effectively zero.  Therefore, we make $\boldsymbol{\Sigma}$ the $r \times r$ matrix that contains all $r$ values greater than some small, positive $\epsilon$.

Substituting Eq.~\eqref{eq:SVDUjm1} into Eq.~\eqref{eq:expAmatU} we get
\[ \mathbf{Y}_+ = e^{\mathbf{A} \Delta t}  \mathbf{U}\boldsymbol{ \Sigma}\mathbf{ V}^*.\]
Rearranging this equation gives
\begin{equation} \label{eq:tildeA}
\mathbf{U}^* \mathbf{Y}_+ \mathbf{V} \boldsymbol{\Sigma }^{-1} = \mathbf{U}^*
 e^{\mathbf{A} \Delta t}  \mathbf{U} \equiv \tilde{\mathbf{S}}.
\end{equation}

It has been shown \cite{Tu:2013dj} that an eigenvalue of $\tilde{\mathbf{S}}$ is also an eigenvalue of  $e^{\mathbf{A} \Delta t}$. To see this, we consider an eigenvalue $\lambda$ and eigenvector $\mathbf{v}$ of $\tilde{\mathbf{S}}$.  By definition we have $\tilde{\mathbf{S}} \mathbf{v} = \lambda\mathbf{v},$ which is equivalent to  $\mathbf{U}^*
 e^{\mathbf{A} \Delta t}  \mathbf{U} \mathbf{v}= \lambda \mathbf{v}.$ Left multiplying this equation by $\mathbf{U}$ we get \[e^{\mathbf{A} \Delta t}  \mathbf{U} \mathbf{v}= \lambda \mathbf{U}\mathbf{v},\]
 which shows that $\lambda$ is an eigenvalue of $e^{\mathbf{A} \Delta t}$.
 {Additionally, $\hat{\mathbf{v}} = \mathbf{U}\mathbf{v}$ is the associated eigenvector of $e^{\mathbf{A} \Delta t}$ to eigenvalue $\lambda$.}

%

The matrix $\tilde{\mathbf{S}}$  is much smaller than that for $e^{\mathbf{A} \Delta t}$ and we can form $\tilde{\mathbf{S}}$ {\em without forming} $\mathbf{A}$. To create $\tilde{\mathbf{S}}$ we need to know the result of $e^{\mathbf{A} \Delta t}$ applied to an initial condition several times in succession. Then we need to compute the SVD of the data matrix $\mathbf{Y}_{-}$. A direct computation requires $O(M^2N)$ operations, though iterative methods for computing the SVD exist \cite{sanger1994two}. As a comparison,  the QR factorization of $e^{\mathbf{A} \Delta t}$, requires $O(M^3)$ operations. 
Our formulation here requires a constant time step size, though this can be relaxed as shown by Tu, et al.\cite{Tu:2013dj}.
\section{Alpha Eigenvalues of the transport operator} \label{sec:alphas}
We will now demonstrate that we can estimate the alpha eigenvalues of a nuclear system by computing several time steps of a time-dependent transport equation and using the DMD theory presented above to form and compute the eigenvalues of $\tilde{\mathbf{S}}$. We begin by defining the alpha eigenvalue transport problem without delayed neutrons.

Consider the time-dependent transport equation \cite{Bell_Glasstone}
\begin{equation} \label{eq:tdtransport}
\frac{\partial \psi}{\partial t}  = A \psi,
\end{equation}
where $\psi(x,\Omega,E,t)$ is the angular flux at position $x \in \mathbb{R}^3$, in direction $\Omega \in \mathbb{S}_2$, at energy $E$ and time $t$. The transport operator $A$ is given by
\[ A = v(E)(-\Omega \cdot \nabla +-\sigma_\mathrm{t} + \mathcal{S} + \mathcal{F}),\]
\change{\sout{is the transport operator}} with $\mathcal{S}$ and $\mathcal{F}$ the scattering and fission operators:
\begin{equation}\label{eq:scatt}
\mathcal{S} \psi = \int_{4\pi} d\Omega'  \int_{0}^\infty dE' \, \sigmas(\Omega' \rightarrow \Omega, E' \rightarrow E) \psi(x,\Omega',E',t), 
\end{equation}
\begin{equation}\label{eq:fiss}
\mathcal{F} \psi = \frac{\chi(E)}{4\pi}\int_{0}^\infty dE' \, \nu\sigmaf(E') \phi(x,\Omega',E',t), 
\end{equation}
where $\sigmas(\Omega' \rightarrow \Omega, E' \rightarrow E) $ is the double-differential scattering cross-section from direction $\Omega'$ and energy $E'$ to direction $\Omega$ and energy $E$ , $\nu\sigmaf(E')$ is the fission cross-section times the expected number of fission neutrons at energy $E'$, and $\chi(E)$ is the probability of a fission neutron being emitted with energy $E$. The scalar flux $\phi(x,\Omega',E',t)$ is defined as the integral of the angular flux over \change{\sout{all direction} the unit sphere},
\begin{equation}\label{eq:phi}
\phi(x,E,t) = \int_{4\pi} d\Omega\, \psi(x,\Omega,E,t).
\end{equation}

Above, we used a continuous  formulation of the transport problem.  For our calculations later, we will use a discretized transport equation using the multigroup method \cite{Bell_Glasstone} in energy, discrete ordinates in angle, and a spatial discretization.  In this case the time-dependent transport equation can be written as a system of differential equations
\begin{equation} \label{eq:tdtransport_dsic}
\frac{\partial \Psi}{\partial t}  = \mathbf{A} \Psi,
\end{equation}
where $\Psi$ is a vector, and $\mathbf{A}$ is a matrix that represents the discrete transport operator.

To define alpha eigenvalues and eigenfunctions consider a solution of the form $\hat\psi(x,\Omega,E) e^{\alpha t}$, which, using \qref{eq:tdtransport}, leads to the relation
\[
\alpha\hat\psi = A \hat\psi.
\]
The values of $\alpha$ where this relation holds are called $\alpha$ eigenvalues and $\hat\psi$ are the alpha eigenfunctions. In discrete form the alpha eigenvalue problem is 
\[
\mA \hat\psi = \alpha \hat\psi,
\]
where $\Psi$ has the form $\hat{\psi} e^{\alpha t}$.
In general the eigenvalues of the discrete problem are not the same as those for the continuous problem due to discretization\change{\sout{ error}}.  From here on, we consider the discrete problem.

In the alpha eigenvalue problem, we are interested in the eigenvalues of $\mA$.  We can use the DMD decomposition to form the operator $\tilde{\mathbf{S}}$ and compute its eigenvalues, and as a result, the eigenvalues of $e^{\mA \Delta t}$. To do this we begin with an initial condition and compute the solution at $N$ time steps. Then we can form $\mathbf{Y}_{+}$ and $\mathbf{Y}_{-}$, compute the SVD, and get the eigenvalues of $e^{\mA \Delta t}$.

We need a way to relate the eigenvalues of $e^{\mA \Delta t}$ to the $\alpha$ eigenvalues.  \change{The relationship is if $(\alpha, \mathbf{v})$ is an eigenvalue/eigenvector pair of $\mA$ then $e^{\alpha \Delta t}$ is an eigenvalue of $e^{\mA \Delta t}$ with eigenvector $\mathbf{v}$.} These facts can be seen through the  definition of the matrix exponential. Consider an eigenvalue $\alpha$ with eigenvector $\mathbf{v}$ for the matrix $\mA$. Using the definition of an eigenvector, we can show that
\[ \mA^\ell \mathbf{v} = \mA^{\ell -1} (\alpha \mathbf{v}) = \mA^{\ell-2} (\alpha^2 \mathbf{v}) = \dots = \alpha^\ell \mathbf{v}.\]
The definition of the matrix exponential gives
\begin{align}\label{eq:matExp}
e^{\mA \Delta t} \mathbf{v} &= \left( \sum_{\ell=0}^\infty \frac{\Delta t^\ell}{\ell !} \mA^\ell \right) \mathbf{v}  \\ \nonumber 
&= \left( \sum_{\ell=0}^\infty \frac{\Delta t^\ell}{\ell !} \alpha^\ell \right) \mathbf{v}\\ \nonumber
&= e^{\alpha \Delta t} \mathbf{v},
\end{align}
where the last equality uses the Taylor series of the exponential function $\exp(\alpha \Delta t)$ around 0.

Therefore, if $\lambda$ is an eigenvalue of $\tilde{\mathbf{S}}$, and, by construction, an eigenvalue of $e^{\mA \Delta t}$, then
\begin{equation} \label{eq:alphaeig}
\alpha = \frac{\log{\lambda}}{\Delta t}
\end{equation}
is an alpha eigenvalue of the discrete transport operator.

The discussion above suggests the following algorithm for estimating alpha eigenvalues of the discrete transport equation:
\begin{enumerate}
\item Compute $N$ time-dependent steps starting from $\psi_0$ using a numerical method of choice and fixed $\Delta t$.
\item Compute the SVD of the resulting data matrix $\mathbf{Y}_{-}$, and form $\tilde{\mathbf{S}}$.
\item Compute the eigenvalues/eigenvectors of $\tilde{\mathbf{S}}$, and calculate the $\alpha$ eigenvalues from Eq.~\eqref{eq:alphaeig}.
\end{enumerate}
This is an approximate method because the time steps typically will not be computed using the matrix exponential, rather a time integration technique such as the backward Euler method will be used.  The backward Euler algorithm estimates the matrix exponential as 
\[
e^{\mA \Delta t} \approx (\mathbf{I} - \Delta t \mathbf{A})^{-1}.
\]
When we use the DMD method on a data matrix generated by the backward Euler method, we are computing eigenvalues of $(\mathbf{I} - \Delta t \mathbf{A})^{-1}.$ To relate these eigenvalues to the $\alpha$ eigenvalues we use the relation
\[ \alpha \approx \frac{1}{\Delta t}\left(1-\frac{1}{\lambda}\right).\]
This approximation will improve at first order as $\Delta t \rightarrow 0$.

\subsection{Comparison with existing methods}
Standard techniques for computing alpha eigenvalues require solving a series of $k$ eigenvalue problems \cite{hill:1983wj}.  The basis for these methods is that the $\alpha$ eigenvalues  make the equivalent $k$ eigenvalue problem exactly critical when the total cross-section is replaced with $\sigmat(E) + \alpha v(E)^{-1}$. This approach will have problems when $\alpha$ is negative as it can cause negative absorption to arise in lower energy groups.

To address this problem other methods have been developed such as Rayleigh quotient methods \cite{Ortega:2017tw}, the Arnoldi method \cite{Lathouwers:2003fv,Kophazi:2012hv}, and Newton-Krylov methods \cite{Gill:2009wq}.  In these approaches the equations that need to be solved are typically different than those required to solve time dependent transport problems.  The DMD method allows one to get both the time dependent solution and eigenvalues as part of one calculation.  Moreover, \change{\sout{the}} DMD provides an estimate for multiple eigenvalues \change{based on the number of modes excited in the system and the number of steps used. \sout{DMD can also be used as a post-process of a time dependent calculation to find the modes in contributing to the solution evolution at different times.}}

\section{Results for Plutonium Sphere}\label{sec:sphere_results}
 Here we present results for the \change{prompt neutron} solution for a sphere of 99 atom-$\%$ $^{239}$Pu and 1 atom-$\%$ natural carbon using 12 group cross-sections and a simple buckling model for leakage so that we can solve an infinite medium problem. The group structure is detailed in Table \ref{tab:12-group-struct}. We will consider sub and super-critical systems by adjusting the radius of the sphere.  \change{Because we use a simple buckling model for this problem we can directly form the matrix for the transport operator and compute ``exact'' eigenvalues for this model. For DMD the time steps are computed using the backward Euler discretization for time integration. In this and subsequent sections we consider only prompt neutrons.}
 
 \begin{table} \caption{The group edges and centers for the 12-group calculations in this study.} \label{tab:12-group-struct}
 \begin{center}
 \begin{tabular}{l|ll}
 $g$ &  $E_g$ (MeV) & $\bar{E}_g$ (MeV) \\
\hline
0 & 17    &     \\
1 &  13.5  &  15.25  \\
2 & 10     &   11.75 \\
  3 & 6.07     & 8.035\\
  4 & 2.865   &  4.4675\\
  5 & 1.353   & 2.109 \\
  6 & 0.5    &   0.9265\\
  7 & 0.184   &  0.342\\
  8 & 0.0676   & 0.1258\\
  9 & 0.0248   & 0.0462\\
  10 & 0.00912  & 0.01696 \\
  11 & 0.00335  & 0.006235 \\
  12 & 0.000454  & 0.001902\\
\end{tabular}
\end{center}\end{table}

 \subsection{Subcritical Case}
\begin{figure}[htbp]
\begin{center}
\begin{subfigure}{0.8\textwidth}\includegraphics[width = \textwidth]{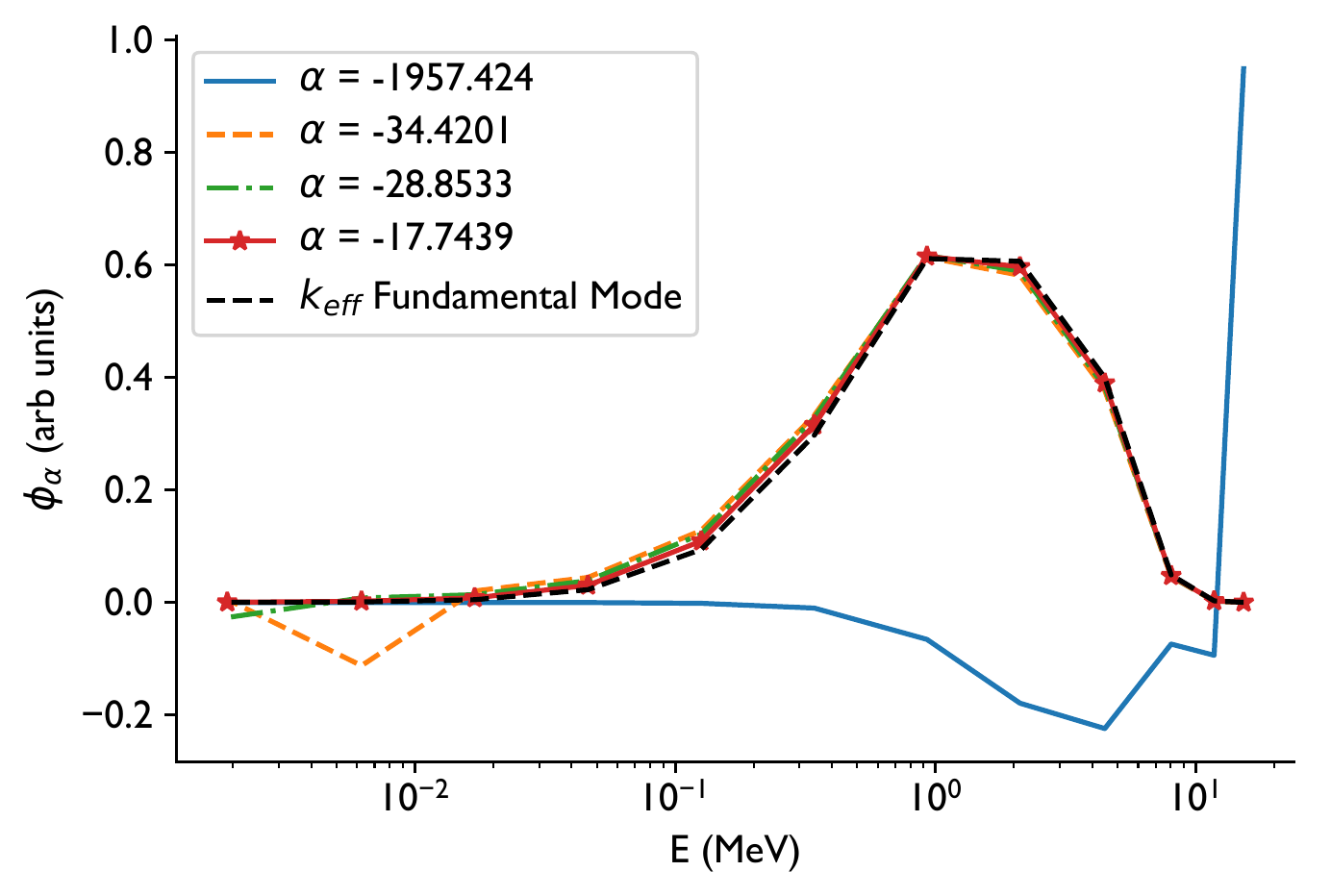}\caption{Subcritical sphere} \label{fig:fundsuba}\end{subfigure}
\begin{subfigure}{0.8\textwidth}\includegraphics[width = \textwidth]{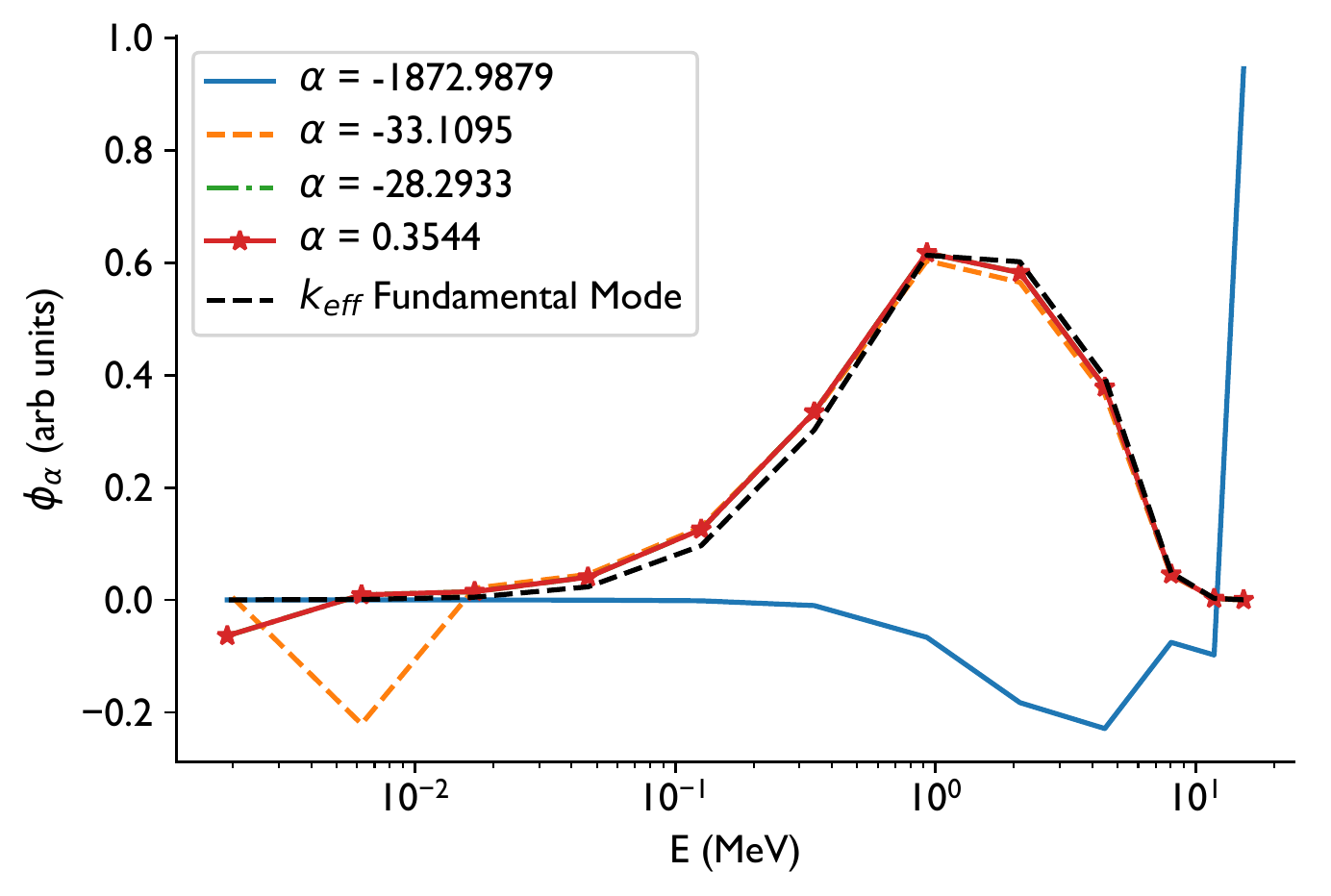}\caption{Supercritical sphere} \label{fig:fundsubb}\end{subfigure}
\caption{Fundmental $k$-eigenmode, and several $\alpha$ eigenmodes for the bare plutonium sphere problem with 12 groups in a subcritical and supercritical configuration. The $\alpha$ eigenvalues have units ($\mu$s$^{-1}$). }
\label{fig:fundsub}
\end{center}
\end{figure}

 We consider a sphere of radius 4.77178 cm with an associated $\keff$ in our model of 0.95000. 
 The fundamental mode for this reactor is shown in Figure \ref{fig:fundsuba} along with several $\alpha$ eigenmodes. The $\alpha$ eigenvalues for this system have a fast decaying mode with a large number of neutrons in the fastest energy group, and the slowest decaying mode closely follows the fundamental mode.
 
\begin{figure}[htbp]
\begin{center}
\includegraphics[width = 0.8\textwidth]{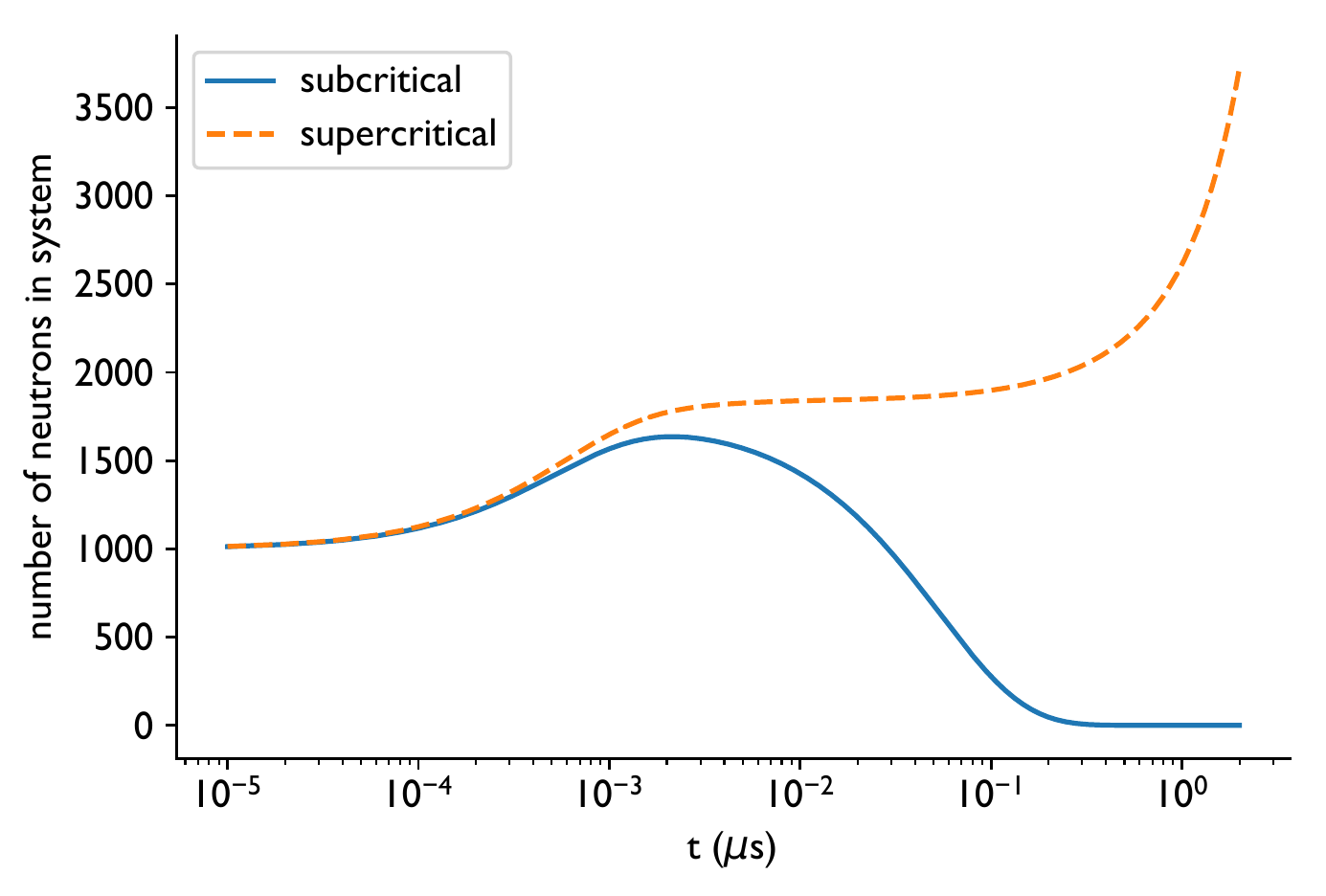}
\caption{The number of neutrons in the plutonium sphere in sub- and supercritical configurations as a function of time. Due to subcritical multiplication the peak number occurs about 0.002 $\mu$s into the simulation of the subcritical configuration. }
\label{fig:td_sub}
\end{center}
\end{figure}

 To test the DMD estimation of $\alpha$ eigenvalues, we run a time-dependent problem where at time zero the system has 1000 neutrons in the energy group corresponding to 14.1 MeV. This is a crude approximation to an experiment where a pulse of DT fusion neutrons irradiates the sphere. The problem is run in time dependent mode out to various final times with uniform time steps, and the time steps are used in the DMD procedure to estimate $\alpha$ eigenvalues.  The $\alpha$ eigenvalues computed by DMD are shown in Table \ref{tab:subcrit} \change{and compared to the exact eigenvalues computed from the matrices generated by the buckling approximation}. The number of neutrons in the system as a function of time is shown in Figure \ref{fig:td_sub}, where one can see that subcritical multiplication is happening in the first 0.002 $\mu$s of the problem. As we argue next,  DMD finds the eigenvalues that are important in the time dependent solution over the time scales considered and that are resolved by the time step size.
 
 From Table \ref{tab:subcrit} we can see that during the phase where subcritical multiplication is occurring (before $t=0.002$ $\mu$s) DMD accurately computes to six digits the $\alpha$ eigenmode that corresponds to a large population of 14.1 MeV neutrons.  This is the mode most excited by the initial condition.  It also accurately computes the eigenvalues with magnitudes larger than 200 to several digits.  However,  we note that the ``dominant'' or slowest decaying eigenmode is not detected by the DMD algorithm, indicating that its contribution at this early time is insignificant or cannot be distinguished from other slowly decaying modes.  This indicates an important phenomenon in time dependent transport: the slowest decaying eigenvalue may not be important in a given problem.
 
 As we look at simulations run to later time, more eigenvalues are identified using DMD.  Running the simulation to intermediate times, 0.02 and 0.2 $\mu$s, we see that DMD finds all of the eigenvalues in the problem to several digits of accuracy.  In both of these solutions DMD does not find the eigenvalue near $-28.85$ $\mu$s$^{-1}$.  This eigenmode has more neutrons in the  energy ranges in the thermal and epithermal energy ranges relative to the other modes.. Given that this problem has very little thermalization due to the small amount of carbon, this mode is not important at these intermediate times relative to other modes.  
 
 At a much later time, 2 $\mu$s, DMD identifies all of the slowly decaying modes but cannot find the rapidly decaying modes.  This is due to the fact that the larger time steps used make it so that the solution cannot resolve the time scale where these modes are important.  As a result DMD estimates a pair of complex eigenvalues with a real part that does not correspond to an actual eigenvalue. \change{There are versions of DMD that allow variable time steps to be used \cite{Tu:2013dj}, and the use of adaptive time stepping should be investigated in future work in order to estimate the fast and slowly decaying modes.}

 \begin{table} \caption{Alpha eigenvalues ($\mu$s$^{-1}$) for the subcritical sphere computed using DMD using the solution obtained with different values of $\Delta t$ and final times.} \label{tab:subcrit}
 \begin{center}
\begin{tabular}{rrrrr}
\hline
      Exact &   $t_\mathrm{final}$ ($\mu$s) = 0.002 & 0.02 & 0.2 & 2\\
\hline
\hline
   -17.7439 &          &           -17.5504 &          -17.7588 &          -17.7437\\
   -28.8533 &           -24.5669 & & &          -28.8628\\
   -34.4201 &           &           -35.7281 &          -34.1948 &          -34.3999\\
   -48.4269 &          &           -46.6817 &          -48.0231 &          -48.4613\\
   -75.0701 &           &           -75.7798  &          -75.2787 &          -74.9998\\
  -132.352  &           &          -132.183  &         -132.197  &         -132.587\\
  -261.942  &          -262.78  &          -261.974 &         -262.127 &         -260.218\\
  -547.732  &              -531.575   &          -547.719  &         -547.11 &         -585.536\\
  -893.385  &            -893.314    &          -893.399  &         -895.262  &         -763.974 \\
 -1368.92   &            -1335.16    &         -1368.90 &        -1362.45\\
 -1732.99   &             -1721.75   &         -1733.01  &        -1725.84 &        -1708 $\pm 381i$  \\
 -1957.42   &             -1957.42  &         -1957.41  &        -1957.42 \\
\hline
& $\Delta t$ ($\mu$s) = 0.0002 & 0.0002 & 0.001 & 0.01\\
\end{tabular}
\end{center}
\end{table}

\subsection{Supercritical Case}
We consider a sphere of radius 5.029636 cm with an associated prompt $\keff$ in our model of 1.000998; the eigenvectors for this problem are shown in Figure \ref{fig:fundsubb}. We perform the same calculations as performed before on the subcritical sphere.  Table \ref{tab:supcrit} compares the eigenvalues computed with DMD with the eigenvalues computed by solving the equivalent infinite medium problem.  At an early time (0.002 $\mu$s), the DMD computation does not identify the exponentially increasing mode.  Upon inspection of Figure \ref{fig:td_sub}, we see that at this time the supercritical and subcritical systems have neutron populations that are very similar.  The subcritical multiplication observed in the smaller sphere where modes associated with the fusion neutrons contributed to the growth of the neutron population, is also present in this supercritical system. However, there are very few neutrons emitted in the fusion energy range from fission ($\chi_1 \approx 1.37\times 10^{-4}$), so these modes decay away.

As the solution time increases the DMD-estimated eigenvalues agree well with the true values.  This is most evident in the solution computed up to $0.2$ $\mu$s where 11 of 12 eigenvalues are computed accurately to 2 digits. The exponentially growing mode is correctly estimated at later times; for the simulation run to the latest time the eigenvalue is estimated accurately to 6 digits by DMD. At very late times the rapidly decaying modes are not correctly estimated and a complex eigenvalue is estimated, as we saw before in the subcritical case, but this is likely due to the large time step used. 

 \begin{table} \caption{Alpha eigenvalues ($\mu$s$^{-1}$) for the supercritical sphere computed using DMD using the solution obtained with different values of $\Delta t$ and final times.} \label{tab:supcrit}
 \begin{center}
\begin{tabular}{rrrrr}
\hline
      Exact &   $t_\mathrm{final}$ ($\mu$s) = 0.002 & 0.02 & 0.2 & 2\\
\hline
\hline
     0.354439 &            -4.02079 &           0.332366 &          0.354291 &          0.354439 \\
   -28.2933   &             & & &        -28.2932\\
   -33.1095   &             &        -32.8048 &        -33.1151\\
   -46.0832   &            &         -45.3512 &        -45.817 &        -46.0703\\
   -70.7945   &             &         -70.4805 &        -70.9448 &        -70.8261\\
  -124.497    &           &        -124.568 &       -124.38 &       -124.381 \\
  -247.14     &             -247.914 &        -247.127 &       -247.281 &       -248.057 \\
  -521.689    &             -506.467       &        -521.693 &       -521.216 &       -507.684\\
  -853.58     &             -853.733      &        -853.577 &       -855.008\\
 -1309.4      &             -1279.91       &       -1309.4  &      -1305.12 & -1050 $+ 23i$\\
 -1659.02     &              -1649.68       &       -1659.02 &      -1655.16 \\
 -1872.99     &             -1872.99      &       -1872.99  &      -1872.98 &      -2059.43\\
\hline
& $\Delta t$ ($\mu$s) = 0.0002 & 0.0002 & 0.001 & 0.01\\
\end{tabular}
\end{center}
\end{table}

\section{Heterogeneous Media}\label{sec:hetero}
The plutonium sphere example required only computing the solution to infinite media problems.  We will now investigate how the DMD approach to estimating eigenvalues performs on a heterogeneous problems in slab geometry.  Our numerical solutions are computed using the discrete ordinates (S$_N$) method with diamond difference for the spatial discretization and backward Euler for  time integration \cite{LewisMiller}. 

\subsection{Heterogeneous, One-speed Slab Problem}
\begin{figure}
\begin{center}
\begin{tikzpicture}[scale=1.0]
\draw (0,0) -- (9.1,0) -- (9.1,4) -- (0,4) -- (0,0);
\filldraw[fill=blue!40!white, draw=black] (0,0) rectangle (1,4);
\filldraw[fill=blue!40!white, draw=black] (8,0) rectangle (9.1,4);
\filldraw[fill=white, draw=black] (1,0) rectangle (2,4);
\filldraw[fill=white, draw=black] (7,0) rectangle (8,4);
\filldraw[fill=black!50, draw=black] (2,0) rectangle (7,4);
\draw (0,0) node[below] {$x=0$};
\draw (1,0) node[below] {$1$};
\draw (2,0) node[below] {$2$};
\draw (7,0) node[below] {$7$};
\draw (8,0) node[below] {$8$};
\draw (9.1,0) node[below] {$X$};

\draw (0.5,2) node {fuel};
\draw (8.505,2) node {fuel};
\draw (4.505,2) node[rectangle,text=black] {absorber};
\draw (1.5,2) node[rotate=90] {moderator};
\draw (7.505,2) node[rotate=90] {moderator};
\end{tikzpicture}
\end{center}
\caption{Layout for the multiregion slab problem from Kornreich and Parsons \cite{Kornreich:2005dc}.  The total width of the problem $X$ can be either $9$ or $9.1$.}\label{fig:korn}
\end{figure}
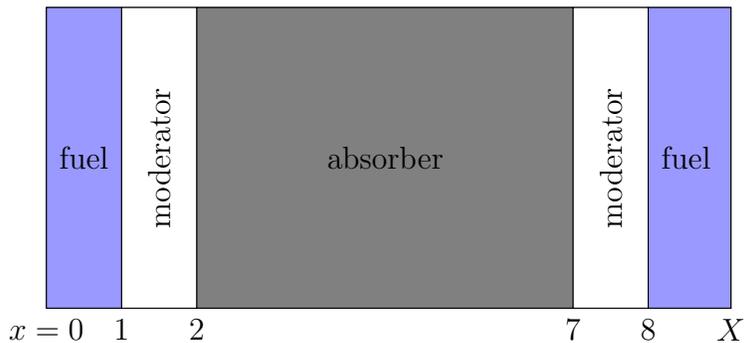

The first heterogeneous problem we solve are based on benchmark problems published by Kornreich and Parsons \cite{Kornreich:2005dc} as solved by the Green's function method (GFM). Their work defines a slab problem for single-speed neutrons (i.e., one group) consisting of an absorber surrounded by a moderator and fuel; see Figure \ref{fig:korn}. They define configurations of this problem that are symmetric and asymmetric, as well as subcritical and supercritical versions.  In the symmetric version of problem the total width of the slab is 9, whereas, in the asymmetric version the width is 9.1. The total cross-section is one throughout the problem and the scattering cross-sections are
\[
\sigmas(x) = \begin{cases}
0.8 & x \in \text{fuel or moderator} \\
0.1 & x \in \mathrm{absorber}.
\end{cases}
\]
The value of $\nu\sigma_\mathrm{f}$ in the fuel is either $0.3$ or $0.7$ for the subcritical and supercritical cases, respectively.

We solve this problem using DMD with 200 cells per mean free path and a 196-angle Gauss-Legendre quadrature set. We use a time step size of $\Delta t = 0.1$ and run the problem for 500 time steps to a time of $t = 50$. For initial conditions we used two approaches: a symmetric initial condition where the solution is non-zero and inwardly directed in the outermost cells in the problem, and a random initial condition. In Table \ref{tab:greens}, results from the DMD calculations are compared with the GFM results.   We use the nomenclature of ``fundamental'' for the alpha eigenvalue that is rightmost in the complex plane to coincide with the published results\change{\sout{, though as we have seen this fundamental mode may not be important in a given calculation}}; the ``second'' eigenvalue in the table is the eigenvalue that is just left of the fundamental eigenvalue in the complex plane. The results in the table show that the DMD results were able to reproduce the GFM eigenvalues within $10^{-5}$ (1 pcm). \change{Except for the second eigenvalue in the symmetric case, all the DMD eigenvalues agreed to greater than 1 pcm precision using both initial conditions.}

\begin{figure}[htbp]
\begin{center}
\begin{subfigure}{0.49\textwidth}\includegraphics[width = \textwidth]{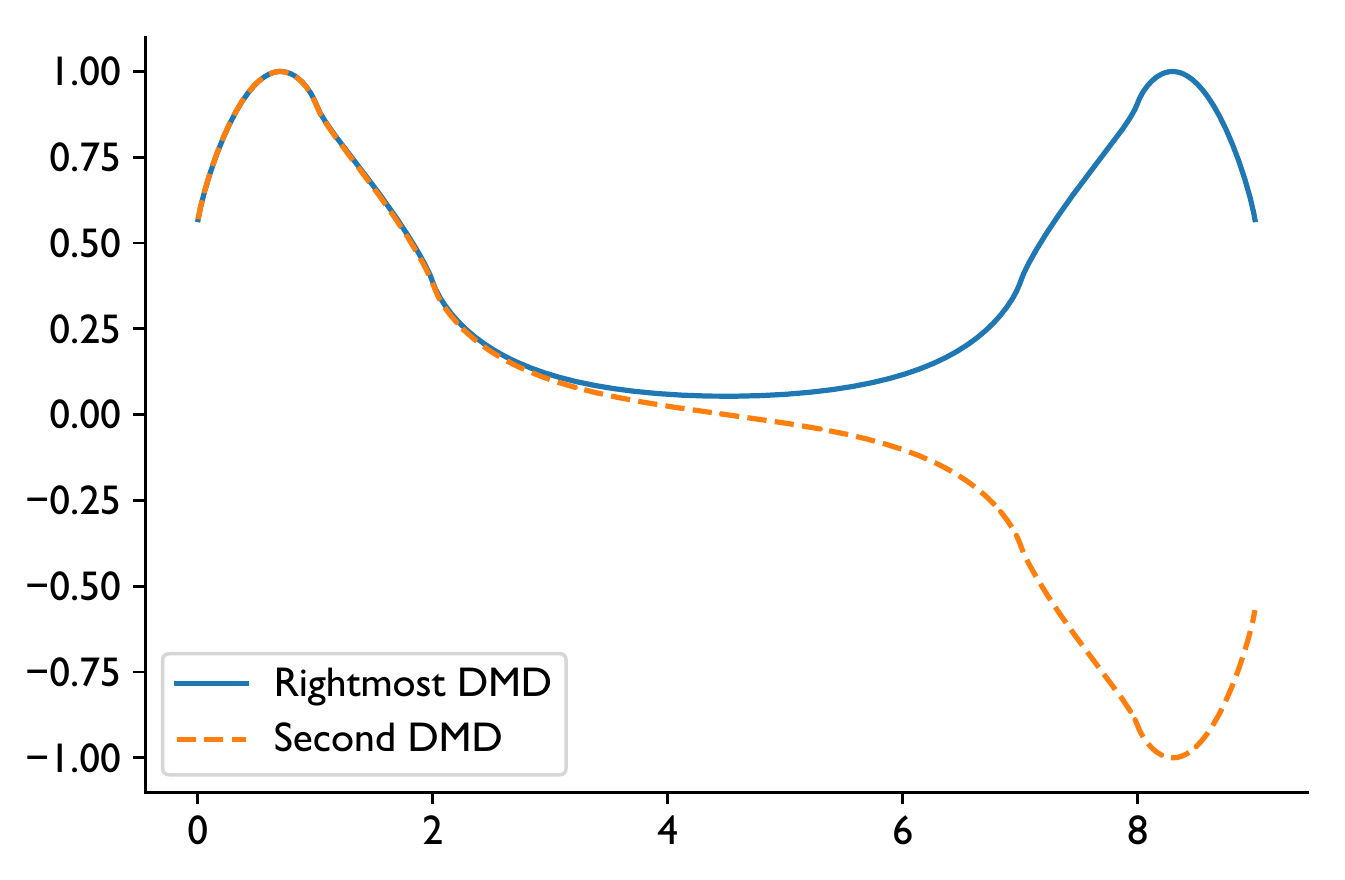}\caption{Symmetric slab with $\nu \sigma_\mathrm{f} = 0.3$}\end{subfigure}
\begin{subfigure}{0.49\textwidth}\includegraphics[width = \textwidth]{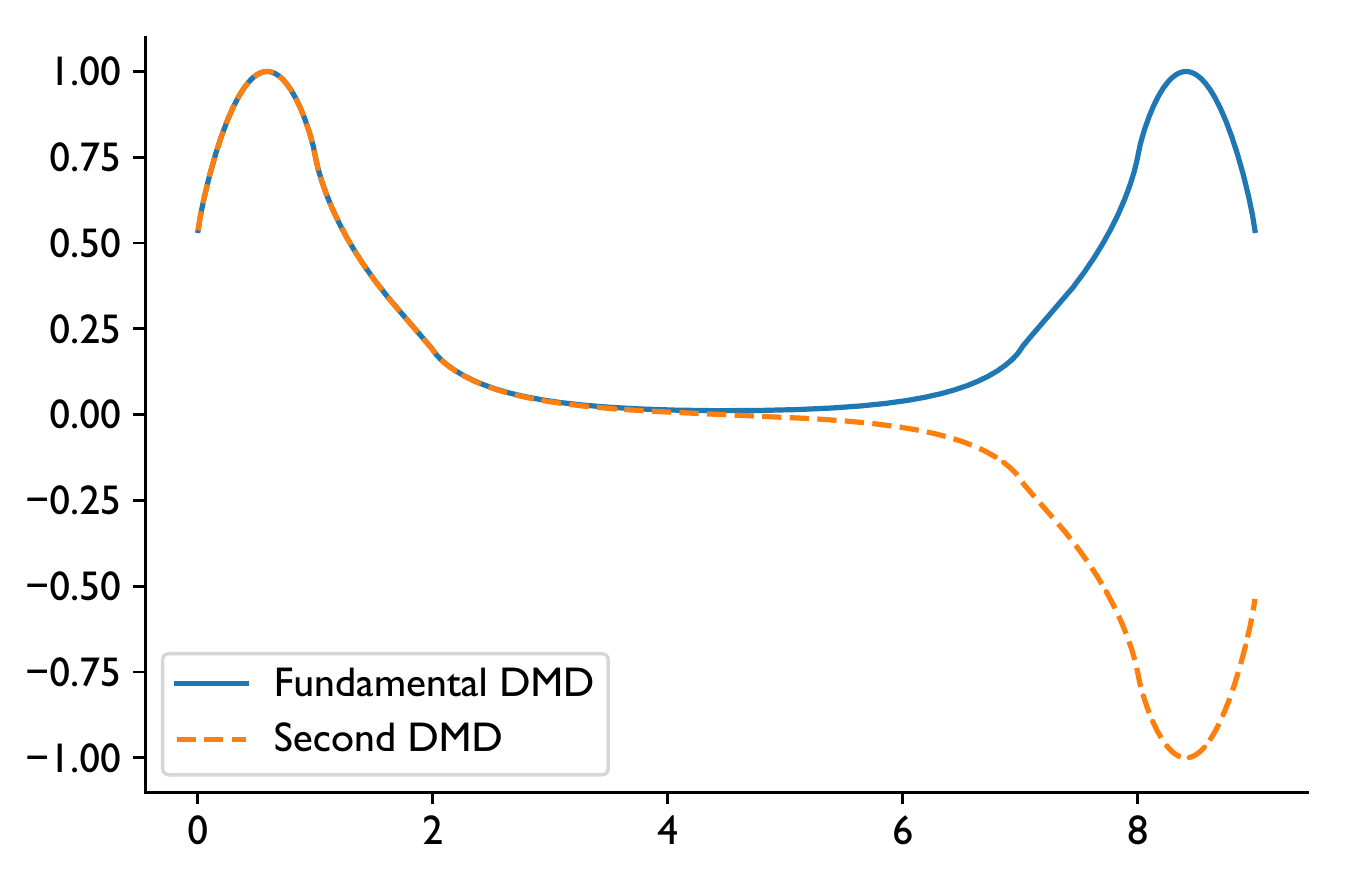}\caption{Symmetric slab with $\nu \sigma_\mathrm{f} = 0.7$}\end{subfigure}
\caption{Fundamental and second eigenmodes for the one group slab problem in the symmetric configurations. }
\label{fig:symm_slab}
\end{center}
\end{figure}

\begin{figure}[htbp]
\begin{center}
\begin{subfigure}{0.49\textwidth}\includegraphics[width = \textwidth]{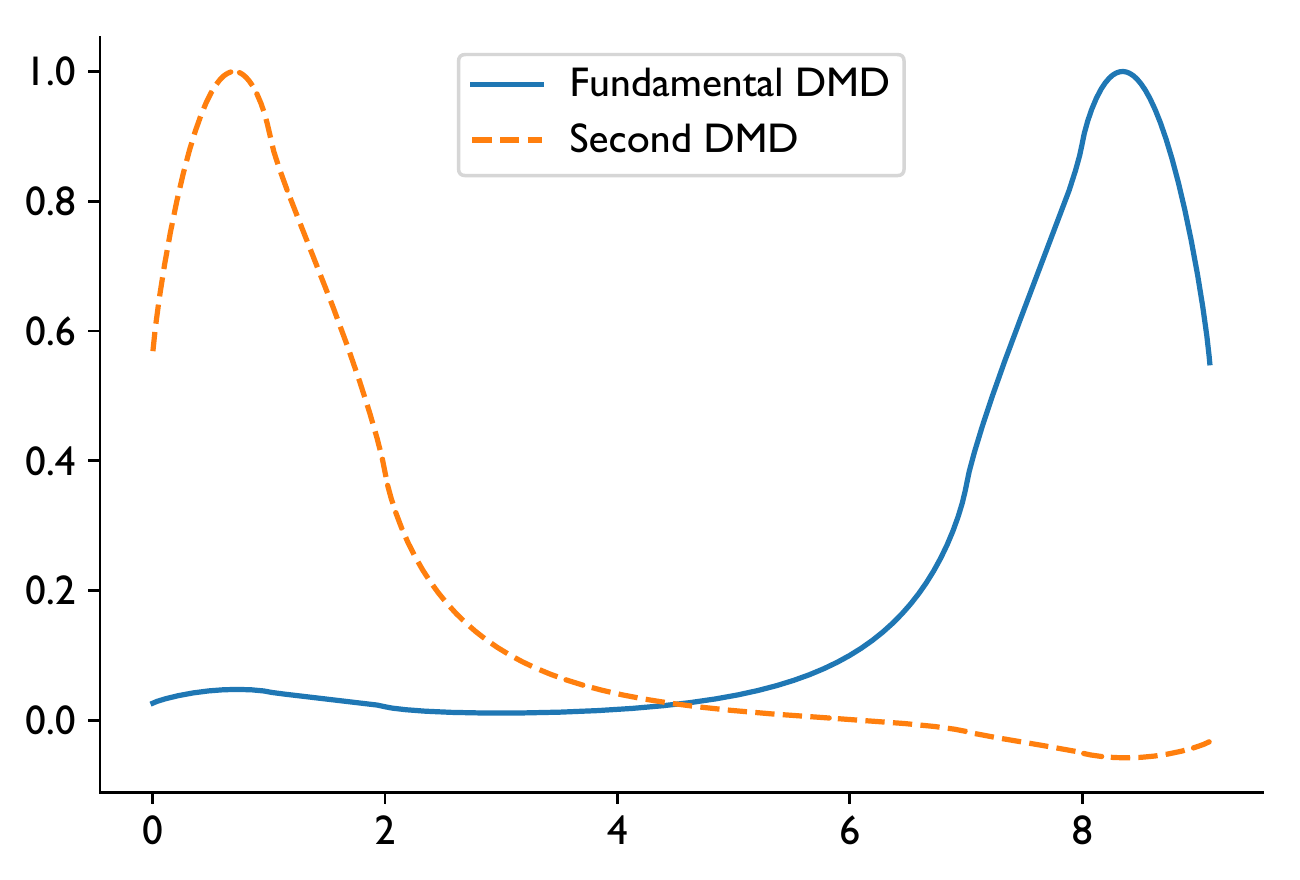}\caption{Asymmetric slab with $\nu \sigma_\mathrm{f} = 0.3$}\end{subfigure}
\begin{subfigure}{0.49\textwidth}\includegraphics[width = \textwidth]{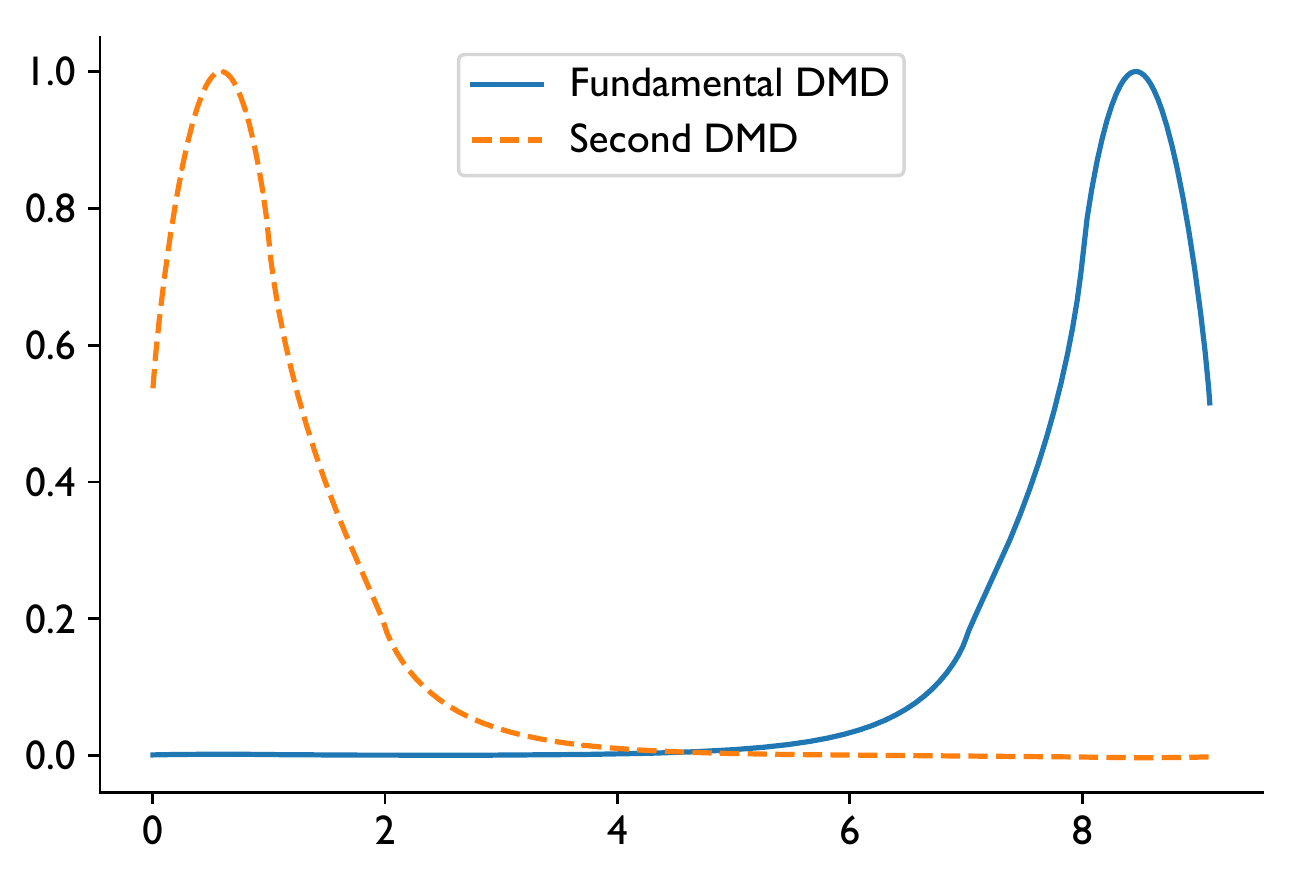}\caption{Asymmetric slab with $\nu \sigma_\mathrm{f} = 0.7$}\end{subfigure}
\caption{Fundamental and second eigenmodes for the one group slab problem in the asymmetric configurations. }
\label{fig:asymm_slab}
\end{center}
\end{figure}

The DMD results in Table \ref{tab:greens} for the fundamental eigenvalue were the same for both initial conditions to six significant digits. We also have found that the eigenvalues found in the solution is  insensitive to the number of time steps used in the DMD procedure, as long as any initial transients have died out (about 5 mean-free times in this problem). Using 400 or 100 time steps in the eigenvalue estimate gave the same eigenvalue estimates to 6 significant digits.  However, the second eigenvalue was not present in the solution for the symmetric initial condition on the symmetric problems.  This is because the second eigenmode is asymmetric in space, and, therefore, this mode is not excited by the symmetric initial condition. The DMD eigenvectors for the four configurations of this problem are shown in Figures \ref{fig:symm_slab} and \ref{fig:asymm_slab}. The fundamental and second eigenvectors match the published plots for the $\nu\sigma_\mathrm{f} = 0.7$ within the width of the lines. 
In the DMD results we found a third, \change{real-valued eigenvalue}, $\alpha = -1.02158875$. \change{\sout{As shown in Figure \ref{fig:asymm_slab} this eigenvalue could be a result of it captures some of the rapidly changing behavior in the fuel and moderator, and this mode could be the result of the angular discretization.} This eigenvalue is part of the continuum spectrum for the transport operator for this problem.  The fact that it is found by DMD is an artifact of the approximations made in the method.}

 \begin{table} \caption{Eigenvalues for the benchmark  as computed via the GFM and the difference between the GFM and DMD estimates in  pcm ($10^{-5}$). } \label{tab:greens}
 \begin{center}
\setlength\extrarowheight{7pt}
{\footnotesize
\begin{tabular}{c|c|llllll}
\hline
Geometry & $\nu\sigma_\mathrm{f}^\mathrm{fuel}$ & Fundamental $\alpha$ (GFM) & $\alpha_\mathrm{GFM} -\alpha_\mathrm{DMD} $  (pcm) & Second $\alpha$ (GFM) & $\alpha_\mathrm{GFM} -\alpha_\mathrm{DMD} $(pcm) \\ \hline \hline
\multirow{2}{*}{Symmetric} 
& 0.3 &  -0.3196537 & 0.639 & -0.3229855 & 0.694 \\
& 0.7 &  -0.006156369 & 0.7711 & -0.006440766 & 0.7724 \\ \hline
\multirow{2}{*}{Asymmetric} 
& 0.3 &  -0.2932468 & 0.535 & -0.3213939 & 0.666 \\
& 0.7  & 0.03759991 & 0.64 & -0.006298843 & 0.7717 \\ \hline
\end{tabular}}
\end{center}
\end{table}


We note that in the original paper by Kornreich and Parsons \cite{Kornreich:2005dc} they give results from the discrete ordinates code {\tt PARTISN} \cite{alcouffe2000partisn} using 96 quadrature points (about half of what we used), and 2000 mesh cells per mean free path (10 times higher resolution than in our case). The {\tt PARTISN} results agreed with the GFM results to within 0.1 pcm using this much finer spatial grid.  Nevertheless, {\tt PARTISN} was not able to estimate the second eigenvalue in the asymmetric cases, whereas the DMD results are as expected.  Furthermore, the Monte Carlo transport code, {\tt MCNP} \cite{briesmeister1986mcnp} was not able to estimate eigenvalues for any of the $\nu \sigma_\mathrm{f} = 0.3$  cases.  Recently, Betzler, et al.\cite{Betzler:2015ey} published Monte Carlo results for these cases using Monte Carlo Markov Transition Rate Matrix Method.

\subsection{Multiregion, 70-group System}
\begin{figure}[htbp]
\begin{center}
\includegraphics[angle=0,width = 0.6\textwidth]{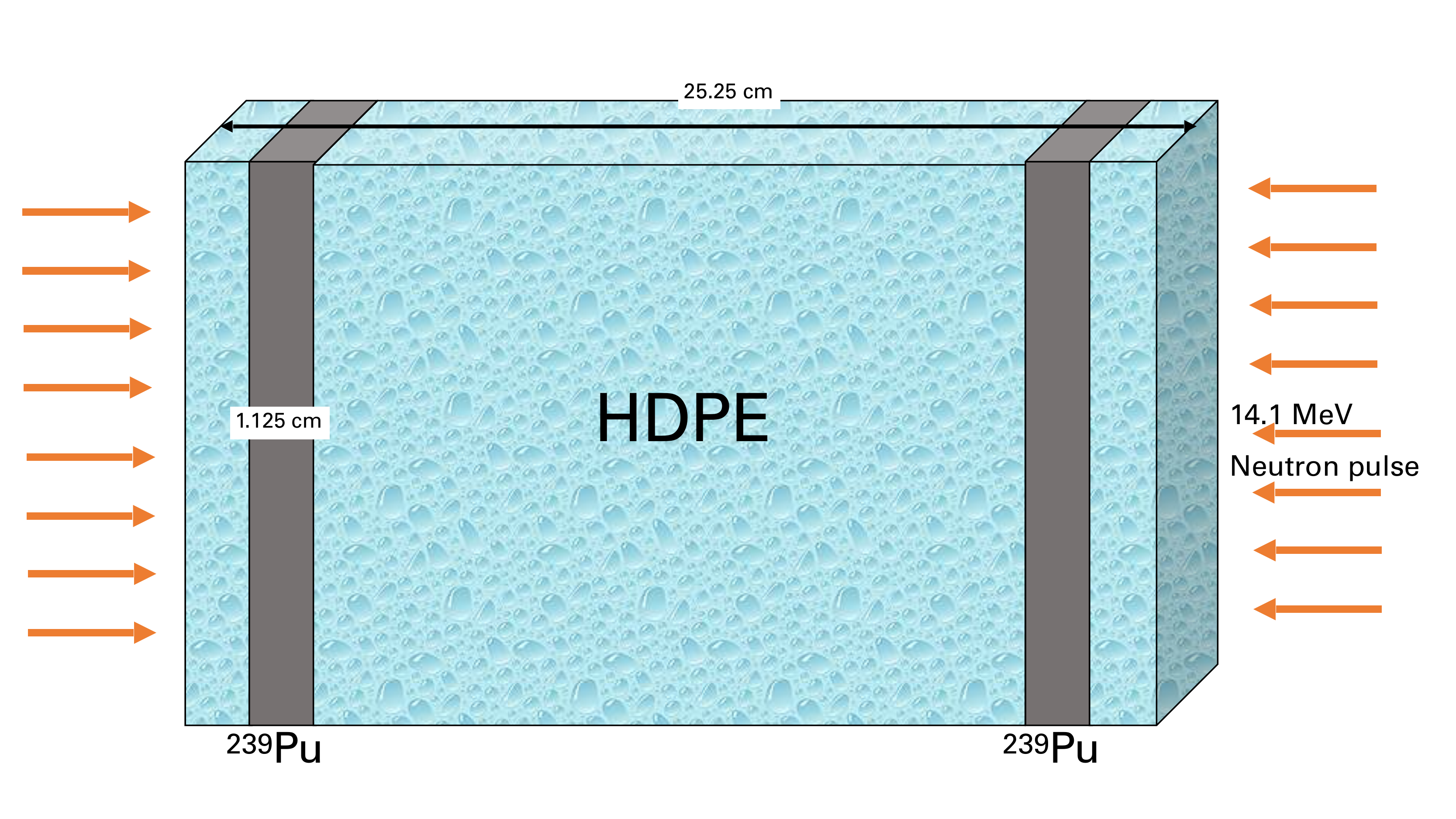}
\caption{Problem layout for the 70-group test problem. }
\label{fig:fundsub_hetero}
\end{center}
\end{figure}

As a final demonstration we solve a problem consisting of two slabs of $^{239}$Pu with high-density polyethylene (HDPE) between them and a reflector of HDPE on the outside.  The initial condition has a pulse of DT fusion neutrons striking the outer surface of the reflector, implemented as the angular flux for each angle directed toward the center being set to 1 in the outermost cell on each side for the initial condition. See Figure \ref{fig:fundsub_hetero} for a schematic of the problem.  The system is subcritical when the fuel regions are each 1.125 cm thick with a resulting $k_\mathrm{eff} \approx 0.97$ and isotropic scattering is assumed. 
The fundamental mode has a large number of thermal neutrons in the middle of the problem as well as a fast peak in the fuel region. 

\begin{figure}[htbp]
\begin{center}
\includegraphics[width = \textwidth]{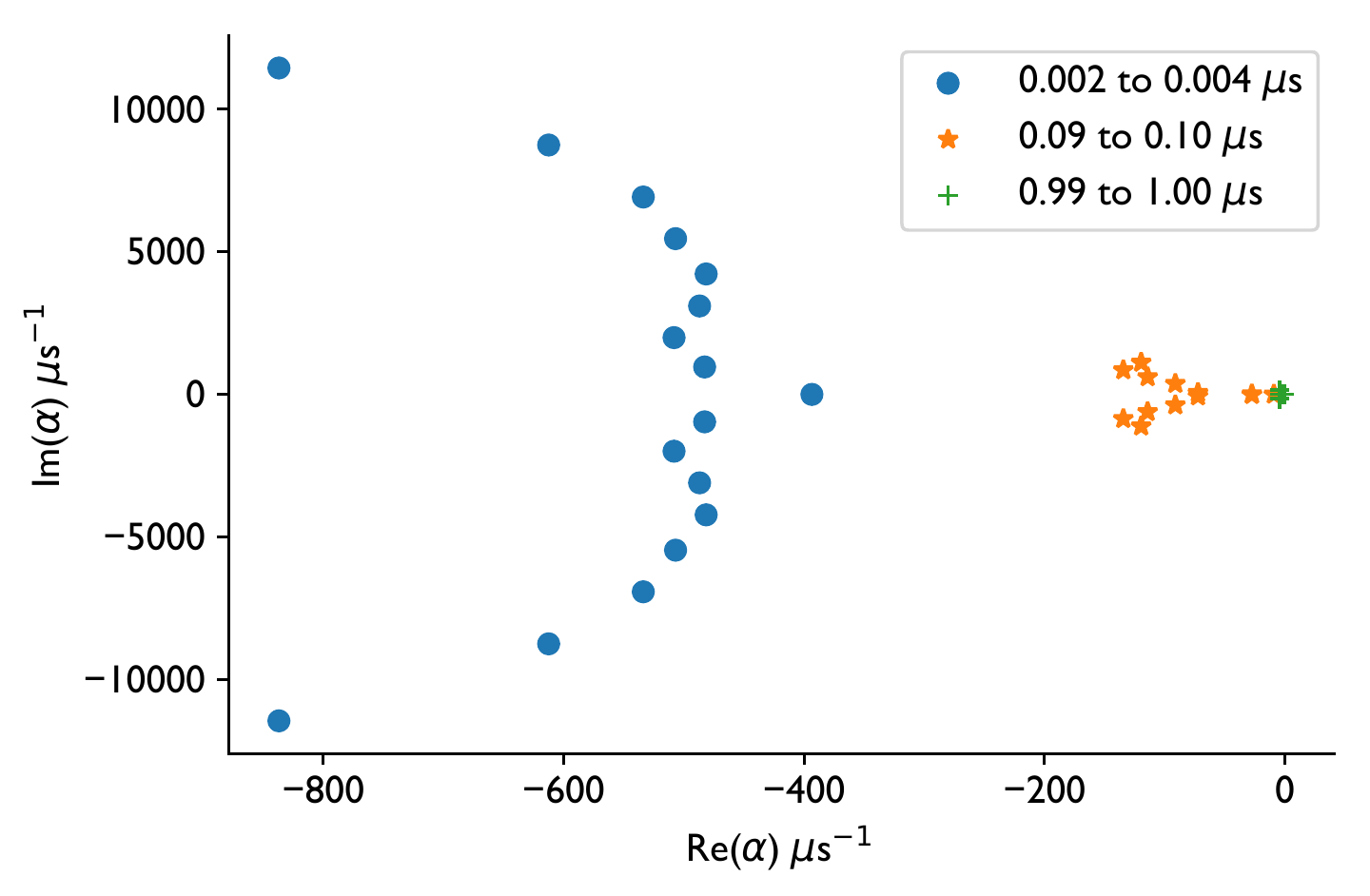}
\caption{$\alpha$ eigenvalues for the 70-group test problem estimated by DMD over three different time intervals. }
\label{fig:fundsub_hetero_eigs}
\end{center}
\end{figure}

Running this problem out to a time of 1 $\mu$s with a time step size of $10^{-4}$ $\mu$s $S_8$ quadrature, and 400 spatial zones, we use DMD to compute eigenvalues present in the solution over three different time windows: 0.002 to 0.004 $\mu$s, 0.09 to 0.1 $\mu$s, and 0.99 to 1 $\mu$s. These eigenvalues are shown in Figure \ref{fig:fundsub_hetero_eigs}. The eigenvalues estimated by DMD at early time (0.002 to 0.004 $\mu$s) have a large imaginary component except for the rightmost value.  As time progresses the imaginary part of the eigenvalues decreases and the real part moves rightward.  This demonstrates a feature of the DMD method: early in time there are many modes present in the solution and the fast decaying ones are governing the solution behavior early in time.  As time goes on, only the slowly decaying modes are present, and DMD finds these later in time. 

\begin{figure}[htbp]
\begin{center}
\begin{subfigure}{0.49\textwidth}\includegraphics[width = \textwidth]{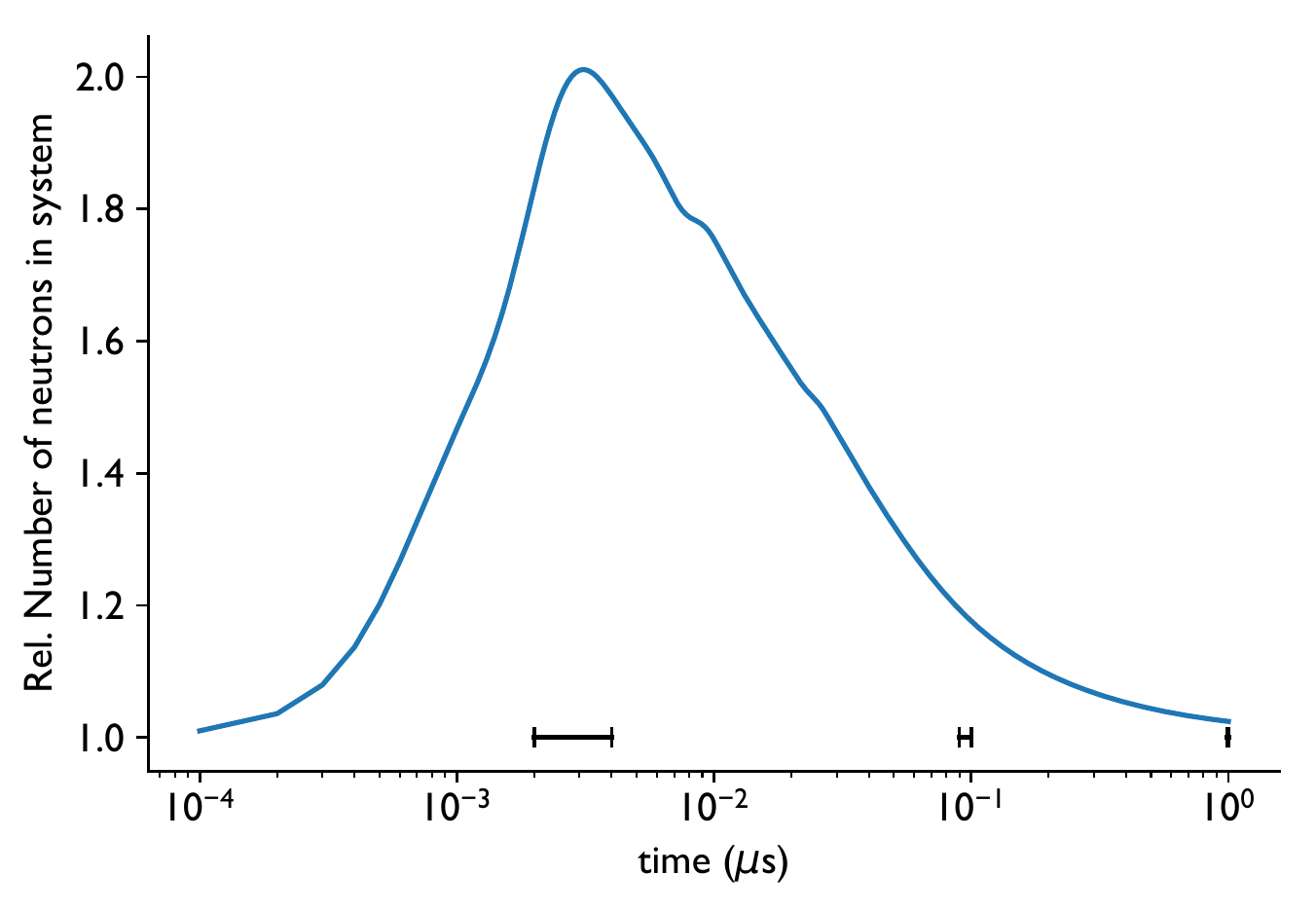}\caption{Neutron population over time}\label{fig:time_dep_window}\end{subfigure}
\begin{subfigure}{0.49\textwidth}\includegraphics[width = \textwidth]{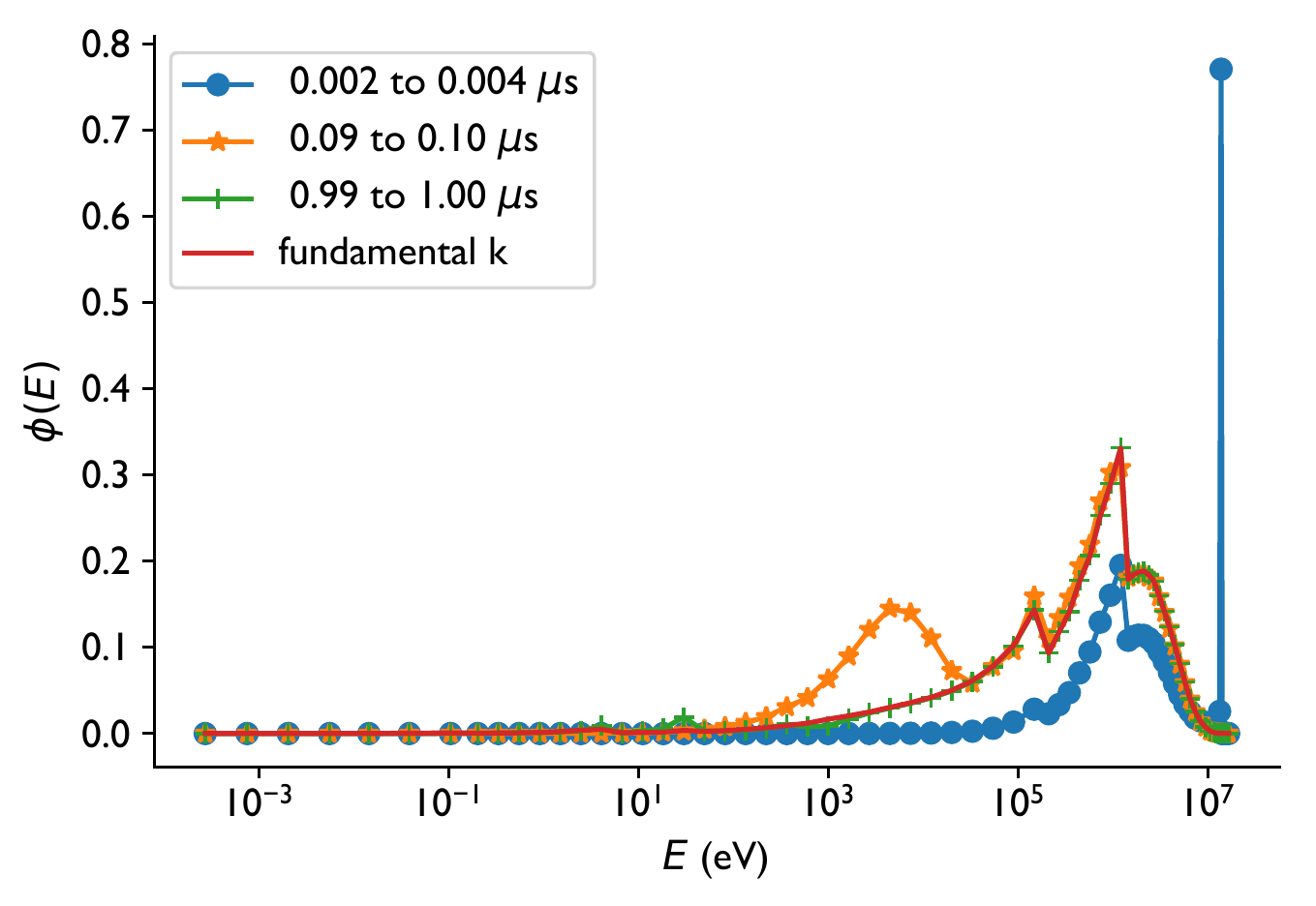}\caption{Midpoint in the outer reflector}\end{subfigure}
\begin{subfigure}{0.49\textwidth}\includegraphics[width = \textwidth]{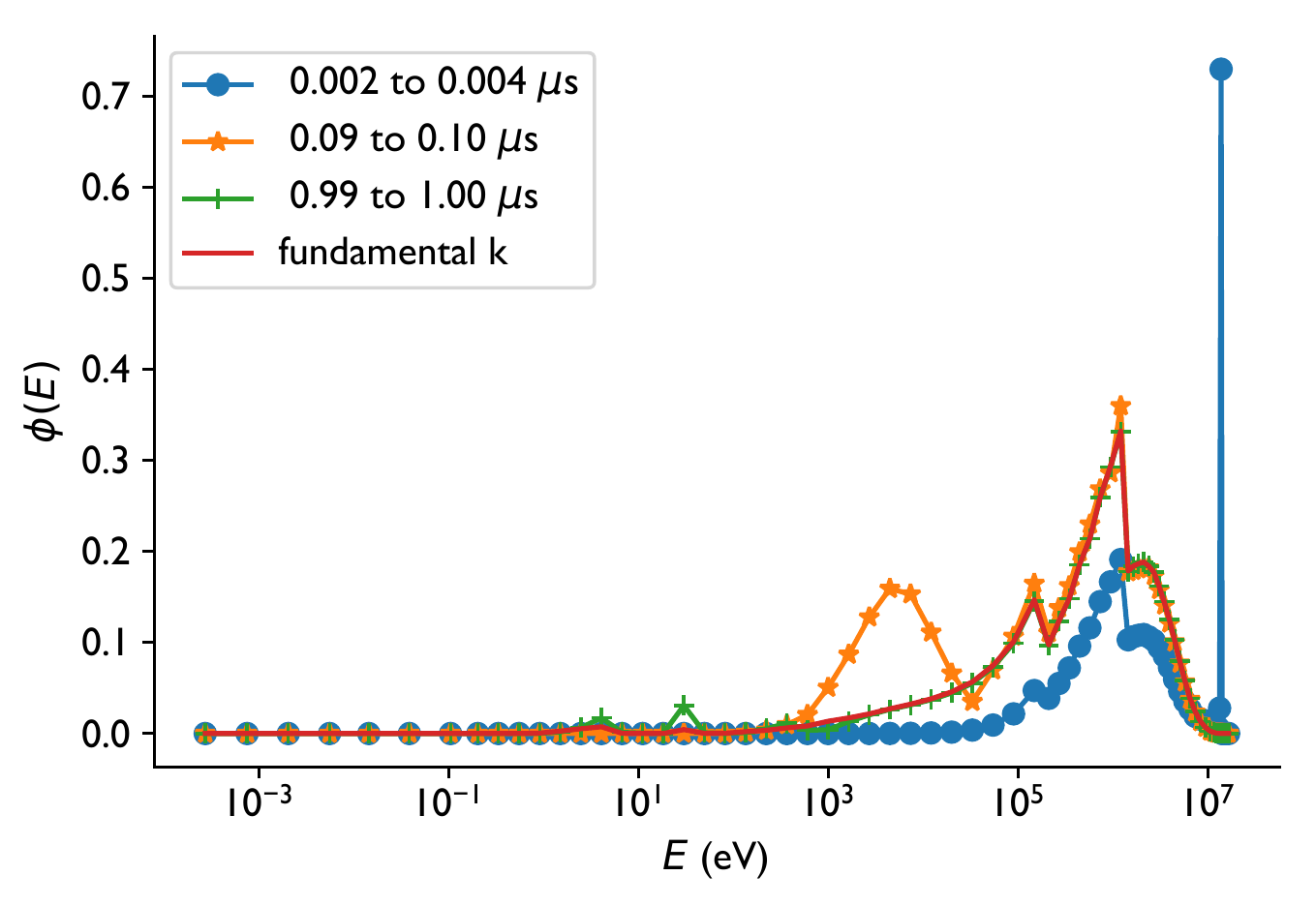}\caption{Midpoint of the fuel}\end{subfigure}
\begin{subfigure}{0.49\textwidth}\includegraphics[width = \textwidth]{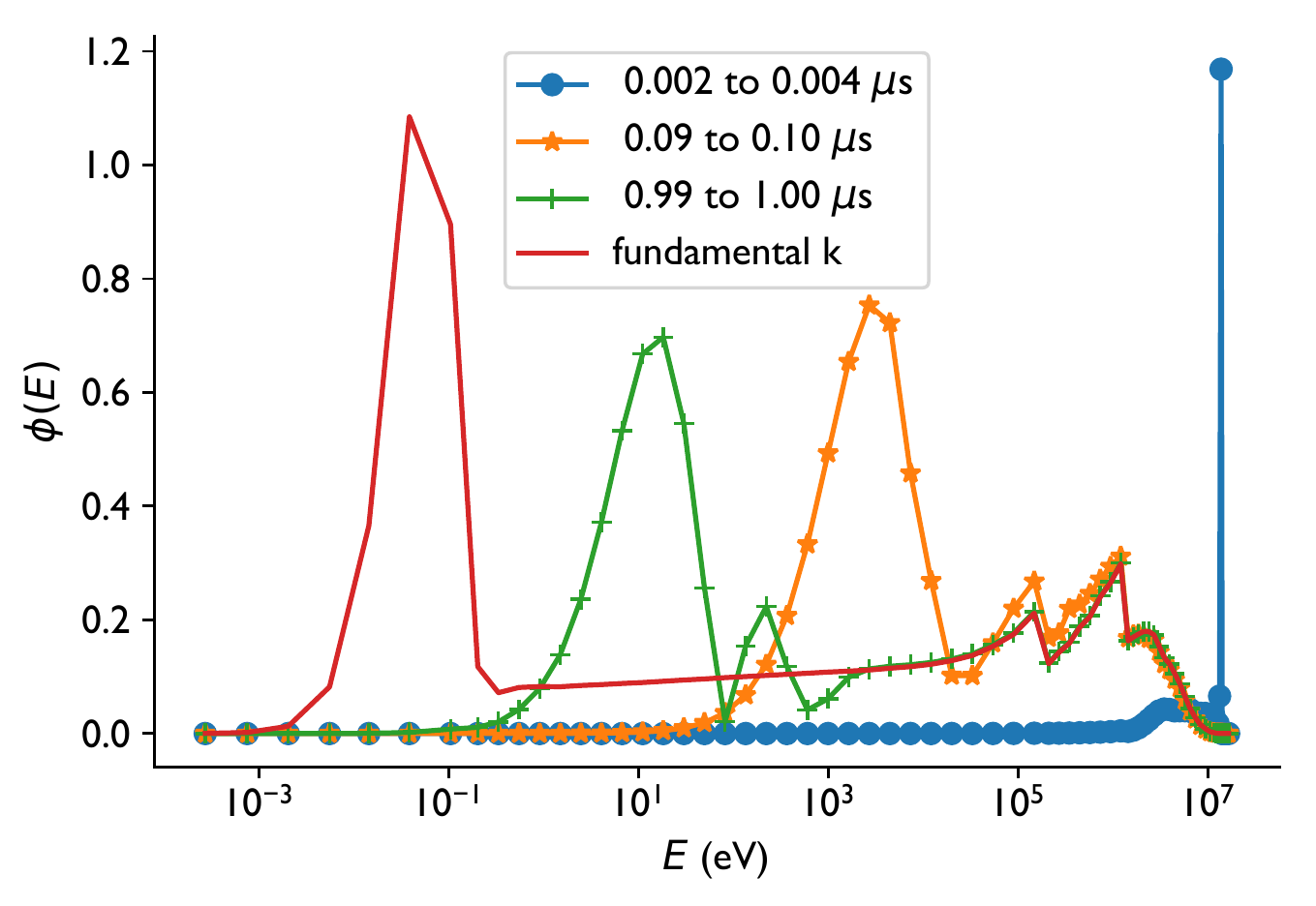}\caption{Problem midpoint}\end{subfigure}
\caption{Neutron population and spectra in the outer reflector, fuel, and moderator \change{averaged} over the three time intervals.  The time intervals are denoted by black lines in (a), and the fundamental $k$-eigenvalue spectra are shown in (b)-(c).}
\label{fig:asymm_slab_time}
\end{center}
\end{figure}

The behavior of the neutron population in time, as well as the three time intervals over which the eigenvalues were estimated is shown in Figure \ref{fig:time_dep_window}. The time interval from 0.002 to 0.004 $\mu$s is during the subcritical multiplication phase of the simulation.  It makes sense that during this phase the slowly decaying modes are not important in the solution. Later in time these slowly decaying modes will dominate because the subcritical multiplication must end at some point given that the system is subcritical and does not have a fixed source.

In Figure \ref{fig:asymm_slab_time} we show the neutron spectrum at several points in space. \change{The spectra shown are computed using time steps from the indicated time ranges.}  From this figure we can see that early in the time the solution is dominated by the presence of 14.1 MeV neutrons, though fission neutrons are present in the fuel and outer reflector. At late times, near 1$\mu$s, the spectrum in the fuel and the reflector is close to the fundamental eigenmode of the $k$-eigenvalue problem. Nevertheless, the central moderator in the problem has not reached the fundamental $k$ eigenmode, as there has not been enough time to fully thermalize the neutrons. Additionally, the eigenvalue for the slowest decaying mode is associated with the travel time of the slowest neutrons crossing the moderator.  This suggests that the problem would need to be run longer to relax to this mode.  Moreover, it indicates that if this system were involved in an experiment, the neutrons produced in the first microsecond would give little information about the spectrum of the $k$ eigenvalue problem. 

\begin{figure}[htbp]
\begin{center}
\begin{subfigure}{0.49\textwidth}\includegraphics[width = \textwidth]{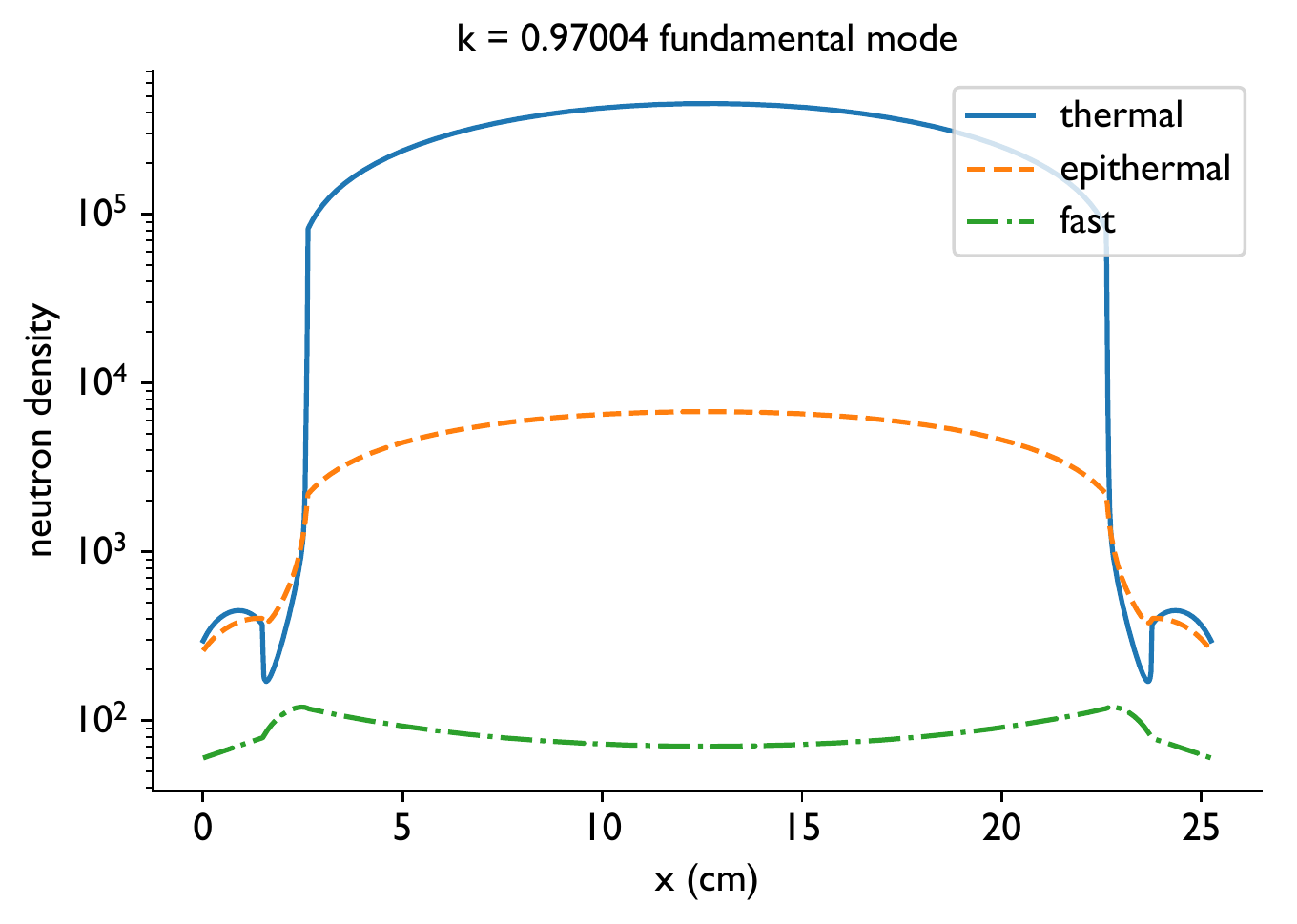}\caption{$k$ fundamental mode}\end{subfigure}
\begin{subfigure}{0.49\textwidth}\includegraphics[width = \textwidth]{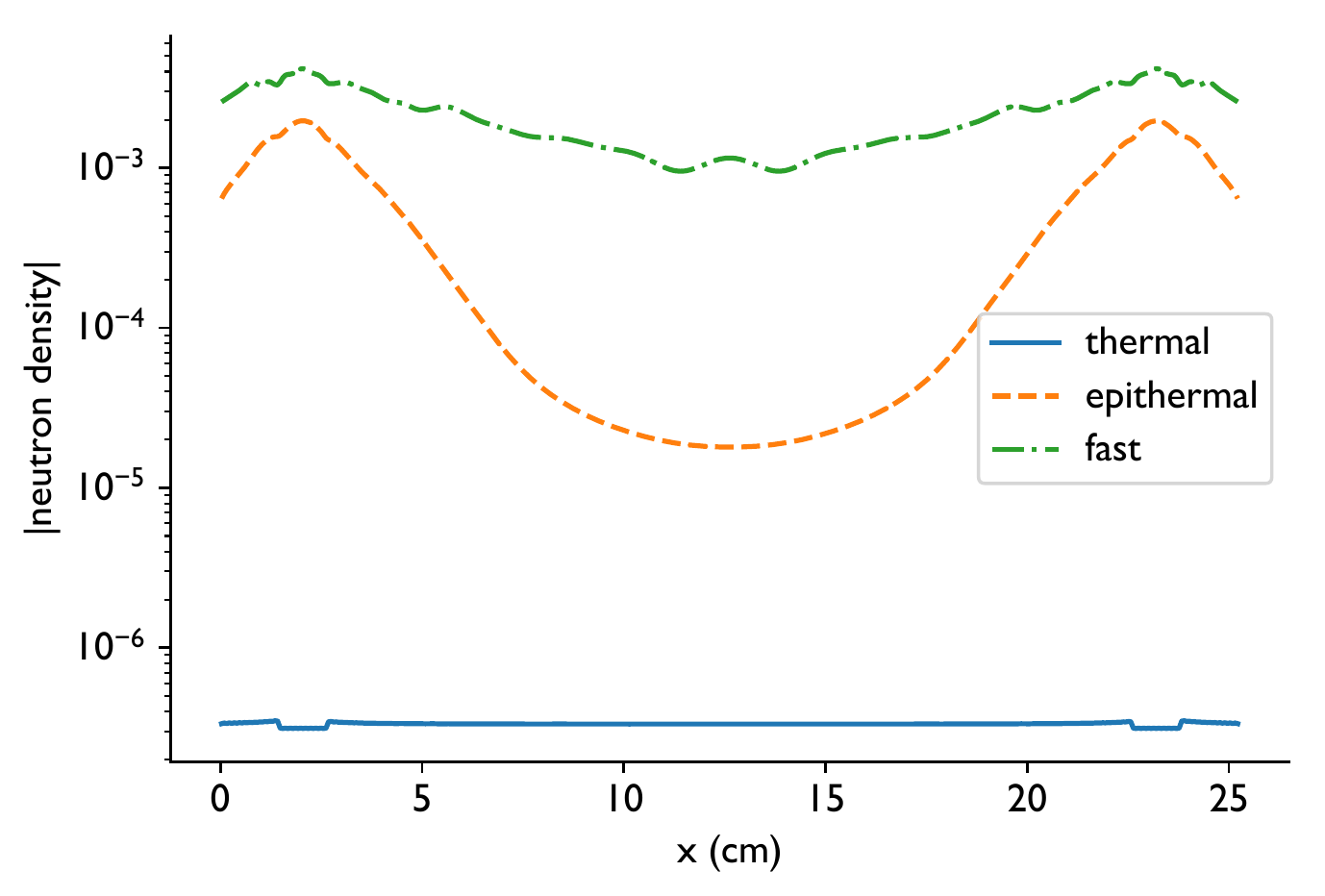}\caption{0.002 to 0.004 $\mu$s, $\alpha = -393.457951\,\mu$s$^{-1}$}\end{subfigure}
\begin{subfigure}{0.49\textwidth}\includegraphics[width = \textwidth]{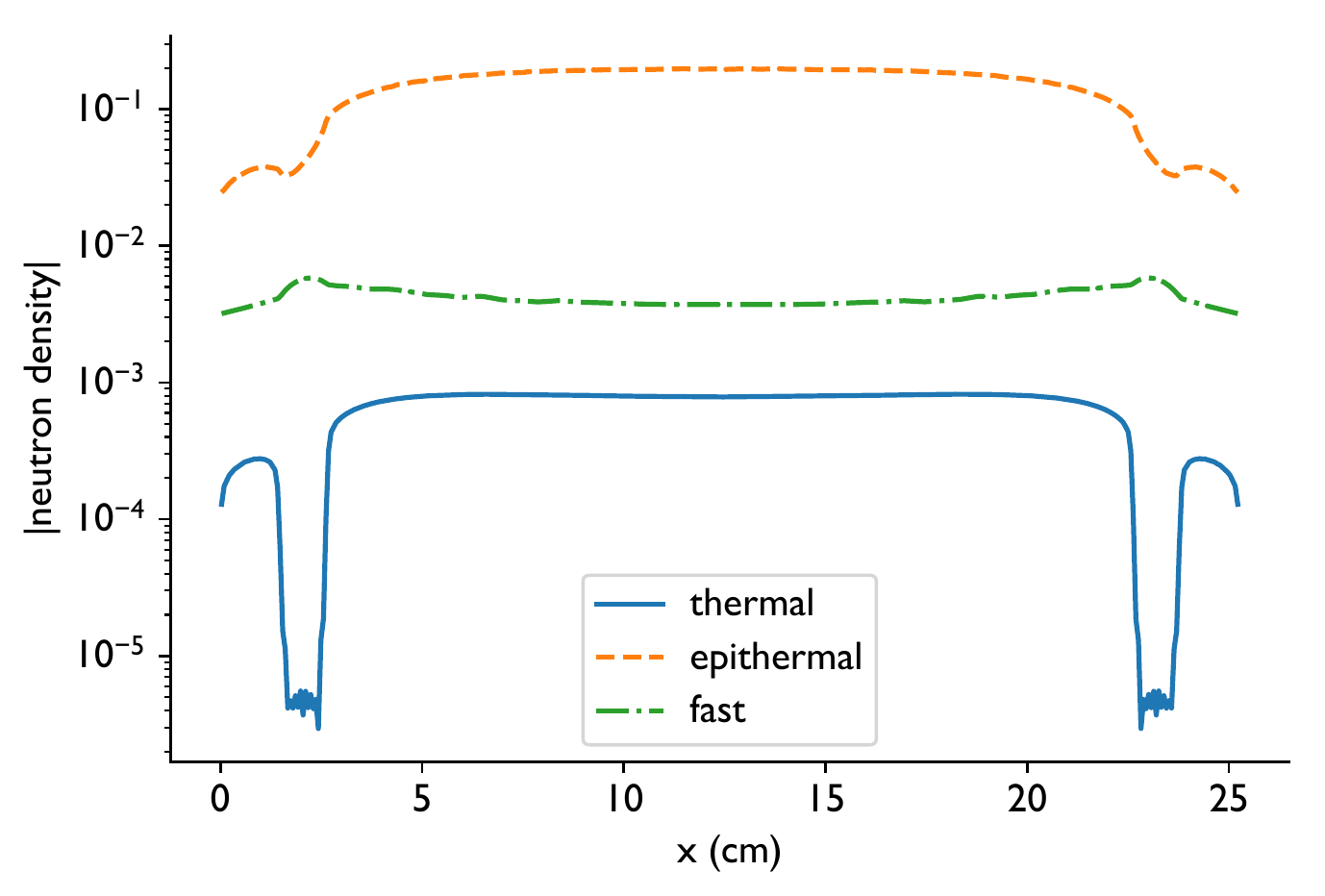}\caption{0.09 to 0.1 $\mu$s, $\alpha = -9.165716\,\mu$s$^{-1}$}\end{subfigure}
\begin{subfigure}{0.49\textwidth}\includegraphics[width = \textwidth]{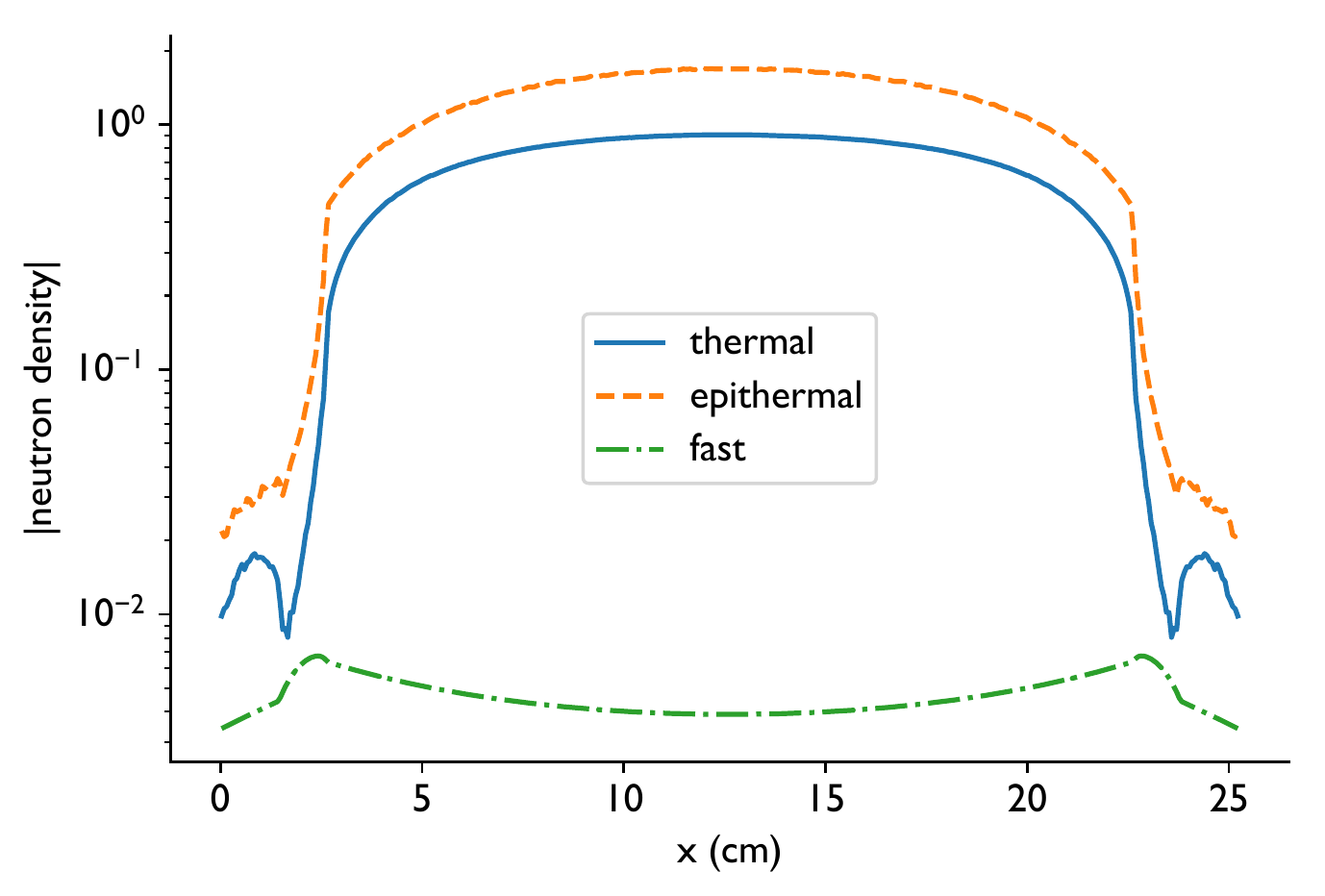}\caption{0.99 to 1 $\mu$s, $\alpha = -0.415305\, \mu$s$^{-1}$}\end{subfigure}
\caption{Spatial distribution of neutrons for the fundamental mode of the $k$ eigenvalue problem, and the eigenvector for the rightmost $\alpha$ eigenvalue as estimated by DMD over  different time intervals. Note that the $\alpha$ eigenvectors are not positive so we plot the absolute value. In this figure thermal neutrons have energy below 5 eV, fast neutrons are above 0.5 MeV, and epithermal neutrons are in between. }
\label{fig:time_fluxes}
\end{center}
\end{figure}

The spatial distribution of neutrons is shown in Figure \ref{fig:time_fluxes}. From this figure we see that at different times the slowest decaying mode that the DMD estimates correspond with the modes that are important to the dynamics during a time interval.  Early in time fast neutrons dominate; these fast neutrons then decay as more thermal neutrons are created from scattering.  Nevertheless, near 1 $\mu$s the \change{\sout{neutrons are not thermalized to the level of the fundamental mode of the $k$ eigenvalue problem.} the neutron density of epithermal neutrons is still larger than the density of thermal neutrons.}

\section{Discussion}\label{sec:discussion}
The dynamic mode decomposition allows for the approximation of the eigenvalues present in a time-dependent transport system from the solution at different times without a separate eigenvalue solve.  The decomposition works for subcritical and critical systems and can give highly accurate (sub-pcm) estimates of eigenvalues.  Our results from a variety of problem types indicate that the method is useful for general estimation of system eigenvalues, especially if one is interested in the modes driving the dynamics over a particular time interval. The problems we presented did not include delayed neutrons, but adding these to the DMD method is straightforward. \change{Because DMD uses the solution from time dependent transport to estimate eigenvalues, the time interval considered and the time step size affect the eigenvalues found. For instance at early times of the simulation there may be different modes present than at later times. DMD will not be able to accurately estimate modes that decay much more quickly than the time step size used to generate the time-dependent solution.} 

We note that DMD can be applied to nonlinear problems in the same fashion as we applied it to the linear problem of neutron transport.  This could be useful for the situation where the neutron population dynamics are nonlinear.  For instance, if we consider a system with negative feedback with regards to temperature, the dynamics of the neutron population would  affect the temperature and the cross-sections of the material.  One could apply DMD to this problem, though the interpretation of the resulting eigenvalues would necessarily be different. Previous work \cite{ROWLEY:2009hb,MEZIC:2013ei}, has shown that the modes computed by DMD will be eigenfunctions of the Koopman operator, and the application of this type of analysis could be fruitful for understanding nuclear systems.


\change{\section{Acknowledgements}
The author would like to thank B.D.\ Lansrud-Lopez, C.D.\ Ahrens, and R.D.\ Baker for helpful discussions during the development of this work.  Also, thanks are in order to S.R.\ Bolding for sharing some {\tt python} code for cross-section processing and infinite media solutions.  LA-UR-18-30110.}

\bibliographystyle{unsrt}
\bibliography{../../RadTran/radtran}

\end{document}